%% file: iccc2022.tex
\documentclass[conference]{IEEEtran}
\IEEEoverridecommandlockouts
\usepackage{amsmath,amsfonts}
\usepackage{amssymb, amsthm}
\usepackage{graphicx}

\usepackage[noend]{algpseudocode}
\usepackage{algorithm}
\usepackage{algorithmicx}

\newcommand*{\algrule}[1][\algorithmicindent]{\hspace*{.5em}\vrule\vrule
width 0pt height \baselineskip depth .25\baselineskip\hspace*{\dimexpr#1-.5em}}

\makeatletter
\newcount\ALG@printindent@tempcnta
\def\ALG@printindent{%
    \ifnum \theALG@nested>0
    \ifx\ALG@text\ALG@x@notext
    \else
    \unskip
    \ALG@printindent@tempcnta=1
    \loop
    \algrule[\csname ALG@ind@\the\ALG@printindent@tempcnta\endcsname]%
    \advance \ALG@printindent@tempcnta 1
    \ifnum \ALG@printindent@tempcnta<\numexpr\theALG@nested+1\relax
    \repeat
    \fi
    \fi
}%
\usepackage{etoolbox}
\patchcmd{\ALG@doentity}{\noindent\hskip\ALG@tlm}{\ALG@printindent}{}{\errmessage{failed to patch}}
\makeatother
\AtBeginEnvironment{algorithmic}{\lineskip0pt}

\algtext*{EndWhile}
\algtext*{EndFor}
\algtext*{EndIf}
\algdef{SE}[DOWHILE]{Do}{doWhile}{\algorithmicdo}[1]{\algorithmicwhile\ #1}%
\usepackage{etoolbox}
\makeatletter
\expandafter\patchcmd\csname\string\algorithmic\endcsname{\itemsep\z@}
{\itemsep=0.0ex plus2pt}{}{}
\makeatother

\usepackage{verbatim}
\usepackage{comment}
\usepackage{geometry}
\usepackage{afterpage}

\makeatletter
\let\MYcaption\@makecaption
\makeatother
\usepackage[font=footnotesize]{subcaption}
\makeatletter
\let\@makecaption\MYcaption
\makeatother

\usepackage{booktabs}
\usepackage{color}
\usepackage{microtype}
\usepackage{balance}
\usepackage{cite}
\input{vmr-symbols-vecbold}
\input{standard-macros}

\input{defs}

\renewcommand{\bml}{\ensuremath{\boldsymbol \ell}}

\newcommand{\eg}[1]{\textcolor{green}{\bf[eg: #1]}}

\safemath{\LAMA}{\textrm{LAMA}}
\safemath{\MRT}{\textrm{MRT}}
\safemath{\betamax}{\beta^\text{max}_\setO}
\safemath{\betamaxno}{\beta^\text{max}}
\safemath{\betamin}{\beta^\text{min}_\setO}
\safemath{\betaminno}{\beta^\text{min}}

\safemath{\Nomin}{\No^\textnormal{min}(\beta)}
\safemath{\Nominnobeta}{\No^\text{min}}
\safemath{\Nomax}{\No^\textnormal{max}(\beta)}
\safemath{\Nomaxnobeta}{\No^\textnormal{max}}
\safemath{\EX}{E_\textnormal{x}}
\safemath{\EXP}{\EX^\textnormal{p}}
\safemath{\Eo}{E_0}

\safemath{\tmax}{{t_\textnormal{max}}}
\safemath{\MAP}{\textrm{MAP}}
\safemath{\IO}{\textrm{IO}}
\safemath{\JO}{\textrm{JO}}
\safemath{\Nopost}{N_{0}^\textnormal{post}}
\safemath{\MT}{U}
\safemath{\MR}{B}
\safemath{\Tran}{\textnormal{T}}
\safemath{\Herm}{\textnormal{H}}
\safemath{\row}{\textnormal{r}}
\safemath{\col}{\textnormal{c}}

\safemath{\NT}{N_\textnormal{T}}
\safemath{\DSNR}{\delta \textnormal{SNR}}
\safemath{\betaMOR}{\beta^{\star}}

\title{A Hybrid Millimeter-wave Channel Simulator for Joint Communication and Localization}

\author{
\IEEEauthorblockN{Junquan Deng} \\
\IEEEauthorblockA{
\small \textit{National University of Defense Technology, China, Email:  {jqdeng@nudt.edu.cn}}\\
\thanks{This work was supported  by the National Science Foundation of China under grant 61901497, 62131005, 62231012 and Research Project of National University of Defense Technology under grant ZK 19-09. The
channel simulator is available at https://github.com/dengjunquan/OmniSIM.}
}
}

\begin{document}
\maketitle
\begin{abstract}
Joint communication and localization~(JCL) is envisioned to be a key feature in future  millimeter-wave~(mmWave) wireless networks for context-aware applications. A map-based channel model considering both site-specific radio environment and statistical channel characteristics is essential to facilitate JCL research and to evaluate the performance of various JCL systems. To this end, this paper presents an open-source hybrid mmWave channel simulator called OmniSIM for site-specific JCL research, which uses digital map, network layout and user trajectories as inputs to predict the channel responses between users and base stations.  A fast shooting-bouncing rays~(FSBR) algorithm combined with Computational Electromagnetic, has been developed to generate channel parameters relevant to JCL, considering mmWave reflection, diffusing, diffraction and scattering.

\end{abstract}


\section{Introduction}
Future wireless networks feature the use of large antenna arrays and wide-band radio access techniques in high radio frequency range, especially in the millimeter-wave~(mmWave) frequencies.
3GPP's 5G New Radio~(NR) in Frequency Range 2~(FR2) and IEEE 802.11ay  are two representative standards for mmWave communications.
Large number of antennas and signal bandwidth adopted in these mmWave networks not only increase the communication capacity,
but also equip the networks with  radio sensing and positioning functions of high accuracies. On the other hand, the built-in localization functionality facilitate various context-aware applications including Internet-of-Things, intelligent transportation system, crowd sensing, assets tracking and advanced radio resource management~(RRM).

The performances of a mmWave joint communication and localization~(JCL) system heavily depends on the site-specific mmWave channel characteristics, including LOS condition, powers of multi-path components~(MPC), angular and temporal distributions of MPCs and Doppler shifts. On the one hand, mmWave communication relies on line-of-sight~(LOS) or prominent reflection paths for beamforming or multiple-input-multiple-output~(MIMO) transmission, and the channel state information~(CSI) needs to be estimated for designing the baseband pre-coder and combiner. On the other hand, various geometrical mmWave localization algorithms also rely on the accurate estimations of the direction-of-arrivals~(DOAs) and/or delays for prominent MPCs. For fingerprinting-based localization methods, channel spatial consistency is important, which describes the smooth variations of channels when the user equipment~(UE) moves in the geographic space.
In order to effectively support the design and evaluation of a mmWave JCL system, accurate network-level channel characterization and modeling considering  massive user equipments~(UEs) are of great importance.

According to how the clusters of MPCs are generated, mmWave channel models can be categorized as stochastic, deterministic and their hybrid ones~\cite{Lim2020,COMST2018}. Stochastic models, including the widely used geometry-based stochastic channel model~(GSCM)~\cite{COST2100,38901,QuaDRiGa2012}, use randomly dropped scatters to generate MPC clusters, they are suitable for evaluating multi-antenna communication performances, but cannot be used directly for designing and evaluating localization techniques. Deterministic channel models are based on detailed site-specific  electromagnetic environment information, using either full-wave solutions such as method of moments~(MoM), finite-difference time domain~(FDTD), or asymptotic technique like geometric optics~(GO) and uniform theory of diffraction~(UTD) to predict the radio channels. As compared, a hybrid model~\cite{Lim2020} uses a simplified geometric description of the propagation environment and a ray-tracing method to generate realistic large-scale spatial channel properties, and adds some stochastic factors to model the small-scale fading effects.

There are several mmWave channel simulators available for academia and industrial research, including  QuaDRiGa~\cite{QuaDRiGa2012}, NYUSIM~\cite{NYU2018}, NIST Q-D Realization~\cite{qd2020}, Altair WinProp~\cite{Winprop} and Remcom Wireless InSite~\cite{alkhateeb2019deepmimo}. QuaDRiGa and NYUSIM are based on GSCM, while Q-D Realization, WinProp and Wireless InSite are based on ray tracing.
QuaDRiGa extends the popular GSCM channel model with new features that allow the generation of channel traces with temporal evolution and scenario transitions. It dose not use an exact geometric representation of the environment but distributes the positions of scatters randomly in the simulated scenario.
NYUSIM provides an accurate rendering of actual channel impulse responses in both delay and spatial spaces based on parameters derived from measurement data.
The open-source NIST Q-D Realization software implements ray tracing to capture the deterministic specular rays, and integrate the deterministic
channel description with stochastic models for diffuse rays. Its ray tracing is based on backtracking algorithm and the method of images and has high computational complexities with a larger number of facets in the simulated scenario, and diffraction and scattering are not implemented.
WinProp is a commercial wireless propagation and radio network planning software, and supports reflection, diffraction and mobile
scattering in the simulation. It employs techniques including the dimension reduction, space partitioning, intelligent ray tracing~(IRT) and dominant path model~(DPM)~\cite{COMST2018} to accelerate the simulation and can be used in a large-scale scenario with thousands of facets.
Wireless Insite is another popular commercial software uses ray tracing coupled with empirical models for a frequency range from 50 MHz to 100 GHz. It employs dimension reduction and ray launching acceleration algorithms, as well as GPU and multi-threaded CPU hardware acceleration.

This paper presents an open-source mmWave network-level hybrid channel simulator~(OmniSIM) for JCL research. It combines a customized deterministic ray-tracing model with stochastic small-scale modeling methods to generated UE-location-dependant channel responses. It models the diffuse reflections by building surfaces using GO, the diffractions by wedges using UTD and the scattering effects by trees using the radiative energy transfer~(RET) function. The core part of OmniSIM is a fast shooting-bouncing rays~(FSBR) algorithm which can find propagation paths between BS and massive UE locations with low computation complexities. OmniSIM is available at https://github.com/dengjunquan/OmniSIM.

\section{MmWave Channel Model}

We consider a generic mmWave orthogonal frequency-division multiplexing~(OFDM) baseband channel
model, assuming a total bandwidth of $W = N \Delta$ with subcarrier spacing $\Delta$, the frequency-domain channel response between a transmitter~(TX)
with $N_T$ antennas and a receiver~(RX) with $N_R$ antennas at subcarrier $n\in \{0,\cdots,N-1\}$ and symbol duration $s \in \{0,\cdots,S-1\}$ is~\cite{Robert2016,Heath2018_ChEst}
\begin{align}
\mb{H}_{n,s}\! =\! \sum_{l=1}^{L} \alpha_l p(sT\!-\!\tau_l) \bm{a}_{r}(\bm{\theta}_{l})\bm{a}_{t}^{\mathsf{H}}(\bm{\phi}_{l})e^{j 2 \pi ( s T  \nu_l-n \Delta \tau_{l})}, \label{eq:ChannelGeneric}
\end{align}
where $L$ denotes the number of propagation paths,  $\alpha_l$ is a complex channel gain for the $l$-th path, $T$ denotes the
symbol duration, $p(\tau)$ is a filter that includes the effects
of pulse-shaping and other low-pass filtering evaluated at $\tau$. Furthermore,
$\bm{a}_{r}(\bm{\theta}) \in \mathbb{C}^{N_{\text{R}}}$ is the RX antenna array response as a function of the {DoA} $\bm{\theta}\in \mathbb{R}^2$ in azimuth and elevation domains, $\bm{a}_{t}(\bm{\phi})\in \mathbb{C}^{N_{\text{T}}}$ is the TX array response as a function of the direction of departure~(DoD) $\bm{\phi}\in \mathbb{R}^2$. Finally, $\tau_l$ is the time of arrival~(ToA) and $\nu_l$ is the Doppler shift, related to the $l$-th path.

The channel models at mmWave differs from conventional sub-6GHz frequencies because of smaller
wavelengths. Diffraction is not significant due to the reduced Fresnel zone, scattering is higher as the wavelengths are comparable to effective roughness of radio reflectors and scatters, and the penetration losses are much larger. A large amount of channel measurement results have confirmed that MPCs for typical outdoor and indoor mmWave channels come with cluster structures in both the delay and angular domains~\cite{Haneda2014,Samimi2016,Cen2018,Charbonnier_TVT_2020}, and the channel power is dominated by LoS and/or specular reflection paths.

In a mmWave JCL system, the baseband RX signal is of the form
\begin{align}
    \bm{y}_{n,s} = \bm{W}_s\mb{H}_{n,s} \bm{x}_{n,s} + \bm{n}_{n,s}, \label{eq:RxSignal}
\end{align}
where  {$\bm{W}_s \in \mathbb{C}^{N_{\text{R}} \times M_{\text{R}}}$ is the RX combiner, using {$M_{\text{R}}\le N_{\text{R}}$} RF chains, $\bm{x}_{n,s} $ is the $s$-th radiated signal vector from the TX antenna array, with $\mathbb{E}\{\Vert \bm{x}_{n,s} \Vert^2\}=P/W$ where $P$ is the average transmit power, and $\bm{n}_{n,s}$ is noise after the combining. For channel estimation and localization purposes, the transmit signals $\bm{x}_{n,s}$ are generally known based on predefined pilots, and the channel $\mb{H}_{n,s}$ can be estimated via $\bm{y}_{n,s}$, $\bm{W}_s$ and a reconstruction algorithm. If the antenna array responses $\bm{a}_{r}(\bm{\theta})$, $\bm{a}_{t}(\bm{\phi})$ are known, it would be possible to estimate the DoAs, delays of dominant paths using MUSIC~\cite{MPCC2018} or other algorithms for positioning purpose. In the case that array responses are unknown, advanced machine learning~{ML} methods can be utilized for localization based on fingerprinting and a channel similarity metric~\cite{SupCon2022}.

\section{OmniSIM Framework}

Different from conventional sub-6GHz BSs, mmWave BSs should be deployed below the rooftops and probably at the same height as the UEs~\cite{BS_height2021}. Notice that, we consider using the vertical-plane-launch~(VPL)~\cite{VPL1998} method to reduce the simulation complexity. Different from a full 3D SBR method, in which rays are launched in the 3D space, the VPL method is a dimension reduction method. OmniSIM first treats vertical building surfaces as 2D segments and generates the rays in the azimuth domain, then  maps the 2D rays to 3D space by considering ground reflections and the possible ceiling reflections.

\begin{figure}
  \centering
  \includegraphics[width=0.95\linewidth]{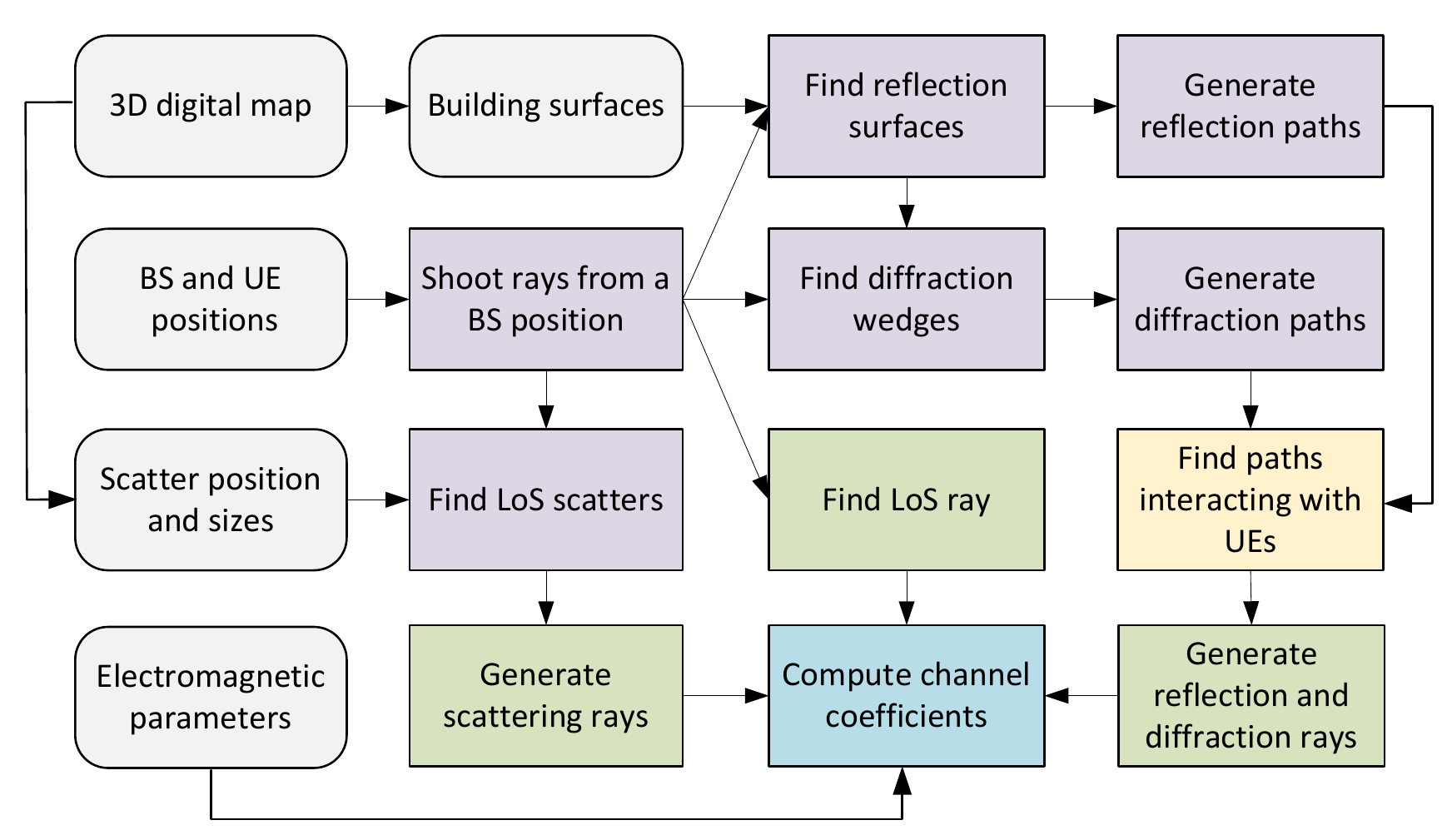}
  \vspace{-0.2cm}	
  \caption{OmniSIM flowchart}\label{fig:flowchart}
\end{figure}

A mmWave channel simulator for JCL should generate realistic MPCs and related path parameters including complex path gains $\{\alpha_l\}$, delays $\{\tau_l\}$, DoAs $\{\bm{\theta}_{l}\}$, DoDs $\{\bm{\phi}_{l}\}$ and Doppler shifts $\{ \nu_l\}$ for massive number of BS-UE links. The direct LoS, specular \& diffuse reflections from building surfaces, diffractions by wedges and scattering by irregular scatters~(e.g. vegetation) need to be appropriately modeled. For this purpose, OmniSIM uses GO, UTD, radiative energy transfer~(RET) models to compute reflection, diffraction and scattering effects. The overall flowchart of OmniSIM is shown in Fig.~\ref{fig:flowchart}, the core part of it is a fast shooting-bouncing rays~(FSBR) method which is given by Algorithm~\ref{Alg:hybrid}. The inputs for OmniSIM are 3D digital map with JavaScript Object Notation~(JSON) files which are constructed from an OpenStreet Map~(OSM), BS positions and orientations, UE positions and orientations, and electromagnetic parameters for various map elements.

OmniSIM first extracts geometric information of building surfaces and scatters~(e.g. trees), then finds the first-order surfaces~(reflectors) and scatters interacted with the direct rays radiated from BS. Each first-order reflector is associated with two diffraction wedges. As diffraction and scattering powers are limited in mmWave frequencies, only the wedges and scatters with LoS condition to both BS and UE at the same time are considered. In comparison, $N_{\rm b}$ bouncing order of reflections with $N_{\rm b} \geq 3$ is considered in OmniSIM. The hybrid propagation paths~(e.g. diffraction-reflection, or scattering-reflection) are not considered as these paths exhibit negligible powers in mmWaves. The proposed FSBR algorithm generates a set of all possible reflection paths $\mathcal{P_{\rm R}}$  with a bouncing order up to $N_{\rm b}$. The function \emph{FindIntersectionSurface}($\mb r$,$\mathcal{W}$) finds the surface first interacted with the ray $\mb r$ and return the index $i$ of the surface $\bm w_i$. If no surface is interacted with the ray, it return an index 0. The function \emph{FindIntersectionPoint}($\mb r$, $\bm w_i$) compute the intersection point among $\mb r$ and $\bm w_i$. If $i=0$, it return a point $\bm p = \bm \infty$ where $\bm \infty$ represents the infinity point. When the number of surfaces is huge, the computational complexity of \emph{FindIntersectionSurface}($\mb r$,$\mathcal{W}$) is high. FSBR avoids calling this function frequently by reusing the pre-computed result via checking whether the previous direct ray has an intersection.

A reflection path is characterized by a sequence of intersection point, while a diffraction or scattering path can be determined by the position of the wedge or the scatter in LoS to both BS and the considered UE position. When generating propagation paths for an UE position $\bm u$, OmniSIM first compute the distances between $\bm u$ and the paths in $\mathcal{P_{\rm R}}$, if the distance is smaller than a predefined threshold, then the path is considered effective for $\bm u$. Collecting all unique reflection paths for $\bm u$ gives a set $\mathcal{P}^{(\bm u)}_{\rm R}$. OmniSIM then will merge $\mathcal{P}^{(\bm u)}_{\rm R}$ with $\mathcal{P}^{(\bm u)}_{\rm D}$ and $\mathcal{P}^{(\bm u)}_{\rm S}$, which are the diffraction paths and scattering paths for $\bm u$.
If $\bm u$ is in LoS to the BS, a LoS BS-UE path should be included.

\begin{algorithm}[t]
\textbf{Input}: Set of surfaces
$\mathcal{W}\! =\! \{\bm w_i\}_{i=1}^{\rm B}$,
BS position $\bm b$,
max bouncing order $N_{\rm b}$,
 predefined angular grid $\mathcal{A}=\{\psi_g\}_{g=1}^{\rm G}$
\begin{algorithmic}[1]
\State Ray index $g \leftarrow 1$,   surface index $i\leftarrow0$, bouncing order $ x \leftarrow 1$, intersection point $\bm p_x \leftarrow \bm\infty$, path set $\mathcal{P_{\rm R}} \leftarrow \varnothing $
\While {$g \leq \rm G$}
\State Shoot a ray $\mb r$  from $\bm b$ to direction $\psi_g$, $ x \leftarrow 1$
\If {$i = 0$ }
\State $i \leftarrow$\emph{FindIntersectionSurface}($\mb r$,   $\mathcal{W}$)
\State $\bm p_x \leftarrow$\emph{FindIntersectionPoint}($\mb r$, $\bm w_i$)
\ElsIf { $i \neq 0$ }
\State $\bm p_x \leftarrow$\emph{FindIntersectionPoint}($\mb r$, $\bm w_i$)
\If {$\bm p_x = \bm\infty$}
\State $i \leftarrow$\emph{FindIntersectionSurface}($\mb r$,   $\mathcal{W}$)
\State $\bm p_x \leftarrow$\emph{FindIntersectionPoint}($\mb r$, $\bm w_i$)
\EndIf
\EndIf
\State $\mathcal{R}=\{\bm b, \bm p_x\}$
\If { $i \neq 0$ }
$I \leftarrow i$
\While {$x \leq N_{\rm b}$ and $\bm p_x \neq \bm\infty$}
\State $\psi \leftarrow $ \emph{ReflectionAngle}($\mb r$, $\bm w_I$, $\bm p_x$)
\State Shoot a ray $\mb r$  from $\bm p_x$ to direction $\psi$
\State $I \leftarrow$\emph{FindIntersectionSurface}($\mb r$,   $\mathcal{W}$)
\State $x \leftarrow x+1$
\State $\bm p_x \leftarrow$\emph{FindIntersectionPoint}($\mb r$, $\bm w_I$)
\State $\mathcal{R}= \mathcal{R}\cup \bm p_x  $
\EndWhile
\EndIf
\State $\mathcal{P_{\rm R}} \leftarrow \mathcal{P_{\rm R}} \cup \mathcal{R} $, $g \leftarrow g+1$ 
\EndWhile
\end{algorithmic}
\textbf{Output}: Set of all possible reflection paths $\mathcal{P_{\rm R}}$
\caption{Fast shooting-bouncing rays~(FSBR) algorithm}
\label{Alg:hybrid}
\end{algorithm}

\subsection{LoS Component}
A LoS cluster has a single ray with the gain calculated by the Friis equation, and the  path coefficient is given by
\begin{align}
\alpha_{\rm LoS} =  \frac{\lambda}{4\pi d} e^{j\frac{2\pi}{\lambda}s},
\end{align}
where $\lambda$ is a wavelength  and $s$ is the distance between TX and RX. Its DoA, DoD can be computed based on the orientations of RX and TX antenna arrays. The delay is $s/c$ with $c$ the speed of light. The Doppler shift is given by $f_{\rm c}\frac{v}{c}\cos\theta_{\rm v}$, with $f_{\rm c}$ the carrier frequency, $v$ the velocity of UE, $\theta_{\rm v}$ the angle between the LoS path and the velocity vector.
\subsection{Specular and Diffuse Rays}\label{sec:ref}

\begin{figure}[b]
  \centering
  \includegraphics[width=0.7\linewidth]{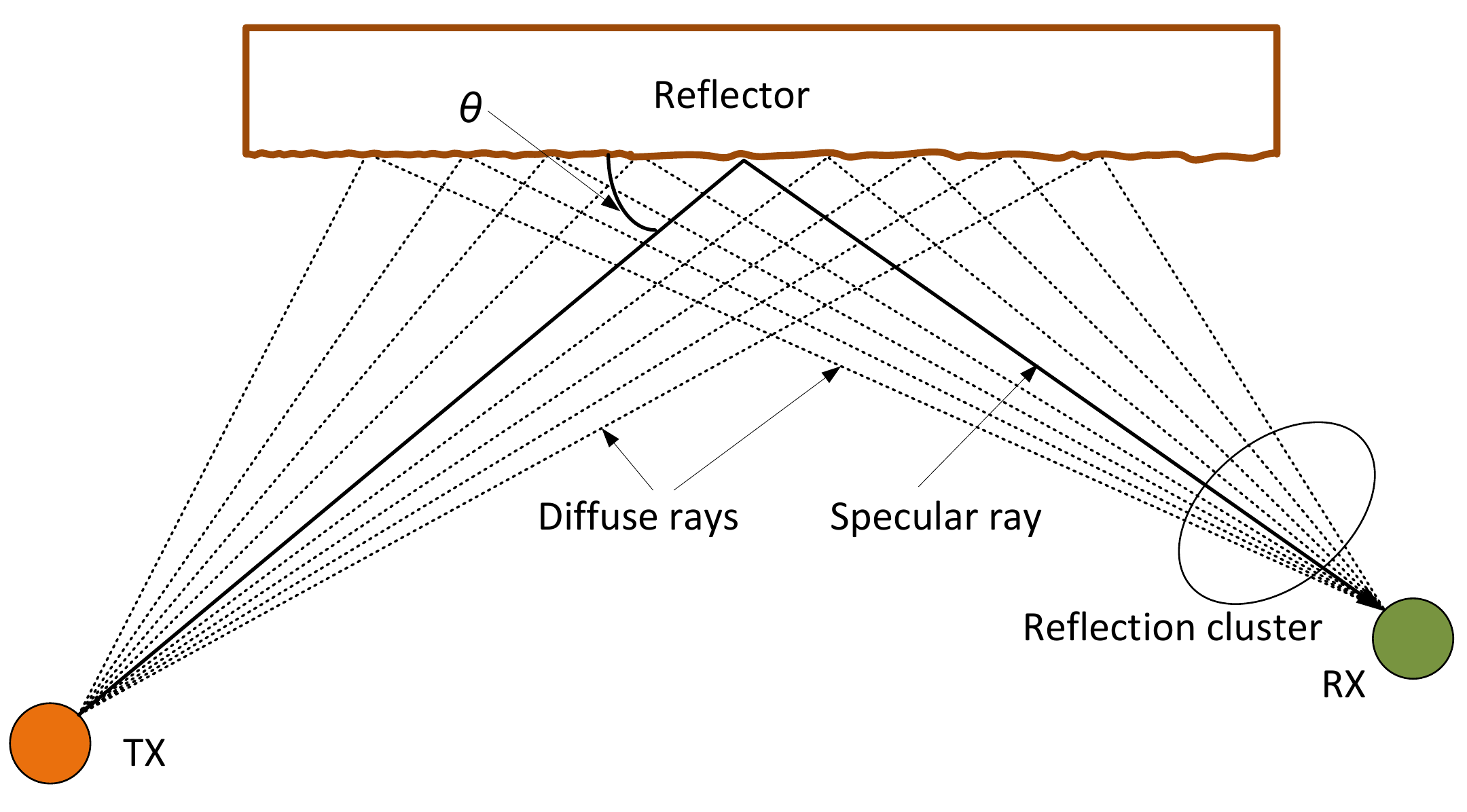}
  \vspace{-0.2cm}	
   \caption{First-order reflection with specular and diffuse rays}\label{fig:reflection}
\end{figure}

A cluster of MPCs relevant to the reflections by a building surface contains a dominant specular ray and diffuse sub-rays  with similar DoAs and delays~\cite{Yaman2021}, as depicted in Fig.~\ref{fig:reflection}.
The Fresnel  reflection coefficients for smooth surfaces can be computed using
\begin{align}
   \Gamma^{\parallel} = \frac{\varepsilon \sin \theta\! -\!\sqrt{\varepsilon \!-\! \cos^2\theta}}{\varepsilon \sin \theta \!+\! \sqrt{\varepsilon \!-\! \cos^2\theta}},  \,\,
   \Gamma^{\perp}  = \frac{                      \sin \theta \!-\!\sqrt{\varepsilon \!-\! \cos^2\theta}}{                      \sin \theta \!+\! \sqrt{\varepsilon \!-\! \cos^2\theta}},
   \label{eq:Reflection}
\end{align}
for vertically and horizontally polarized reflection rays respectively,  where $\theta$ is the grazing angle between reflector surface  and the incidence ray, and $\varepsilon$ is the relative permittivity of the reflector material.
As the wave lengths are small and usually comparable to the roughness of exterior walls, diffuse reflection is inevitable for mmWave signals.
We can calculate a scattering loss factor~\cite{Rappaport1996} due to surface roughness that is used to multiply the result for smooth surface reflection to account for diminished energy in the specular direction
of reflection, which is given by
\begin{align}\label{eq:reflection_loss}
\rho_{\rm S} = \exp \left[ -8 \left( \frac{\pi \sigma_{\rm h}\sin \theta}{\lambda}\right)^2 \right],
\end{align}
where $\sigma_{\rm h}$ is the standard deviation of the zero mean height of the reflection surface.
Measurement results confirmed that the reflection clusters generally
consist of a main peak followed by weaker components
with shorter or longer delays~\cite{Haneda2014} and different DoA and DoD offsets, as shown in Fig.~\ref{fig:reflection}.
To model this phenomenon,  the intra-cluster parameters including the sub-ray offsets, DoA and DoD offsets  w.r.t. the specular ray is modeled using stochastic processes with a pre-defined distribution density function. Specifically, the received power from a reflection surface at the RX is assumed to be contributed  by $n_S$ sub-rays.
The path gain coefficient for a first-order reflection sub-ray is assumed to be
\begin{align}
\alpha_{\rm Relf} = e^{j\zeta} \rho_{\rm S} \Gamma \sqrt{\frac{1}{n_{\rm S}}}  \frac{\lambda}{4\pi d_{\rm T}} ,
\end{align}
where $e^{j\zeta}$ is a random phase,  $d_{\rm T}$ is the total  propagation distance.
The mean  values of the intra-cluster Delay Spreads~(DS) are 3.90 and 12.86~ns~\cite{Haneda2017}, in the in-building and urban scenarios, respectively.
\begin{figure}[t!]
  \centering
  \includegraphics[width=0.8\linewidth]{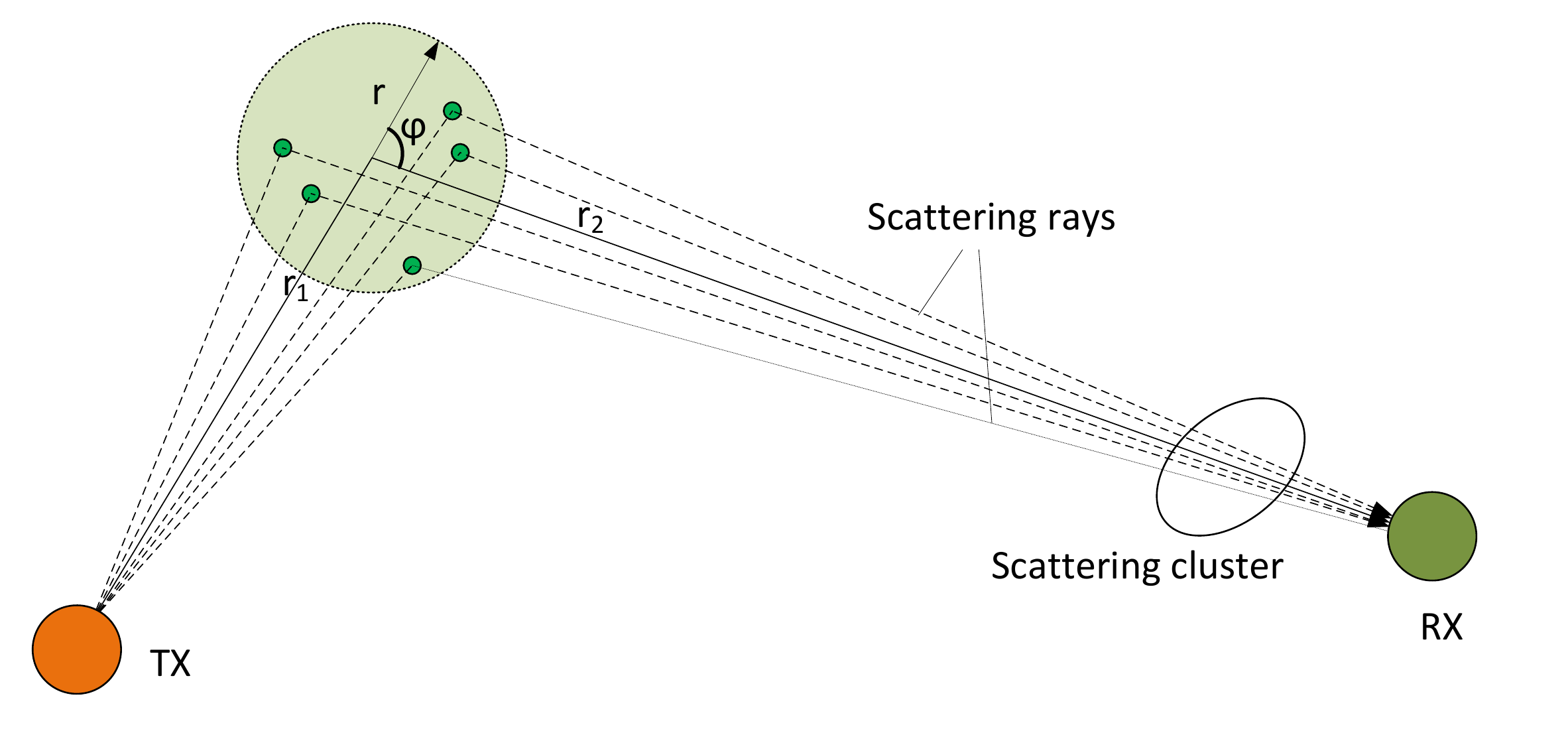}
  \vspace{-0.2cm}	
  \caption{Scattering by a tree}\label{fig:scattering}
\end{figure}

According to measurement results in typical in-building and urban environments~\cite{Haneda2014,Haneda2017}, the intra-cluster delay offsets  generally exhibit an exponential distribution, while the  DoA and DoD offsets in both the azimuth and elevation domain follow a Laplacian distribution.
If we model the intra-cluster delay offset $\tau_o$ as a random variable, its probability density function (PDF) is an exponential function given by
\begin{align}\label{eq:PDF_DS}
  f_{\tau_o}(\tau_o) = \frac{1}{\rm D_S} \exp\left(-\frac{\tau_o}{\rm D_S}\right),
\end{align}
with $\rm D_S$ the intra-cluster delay spread.
Similarly,  the sub-ray DoA or DoD offset~(in azimuth or elevation)  within a cluster can be modeled by a random variable $\psi$, whose PDF is
\begin{align}\label{eq:PDF_AS}
  f_{\psi}(\psi) = \frac{1}{\sqrt{2}\rm A_S} \exp\left(-\frac{\sqrt{2}|\psi|}{\rm A_S}\right),
\end{align}
with $\rm A_S$ the intra-cluster angular spread~(AS).
The measured intra-cluster  AS for departure~(ASD) and AS for arrival~(ASA) of the in-building and urban environments have
mean values of {4.33, 5.94} and {5.82, 15.56} degrees~\cite{Haneda2017}.

For an $n$-th order reflection, OmniSIM uses
\begin{align}
\alpha^{[n]}_{\rm Relf} = e^{j\zeta} \left[\prod_{i=1}^n\rho^{[i]}_{\rm S} \Gamma^{[i]}\right] \sqrt{\frac{1}{n_{\rm S}}}  \frac{\lambda}{4\pi d_{\rm T}} ,
\end{align}
to evaluate the path gain with multiple reflections, where $\rho^{[i]}_{\rm S}$ and $\Gamma^{[i]}$ are the scattering loss factor and reflection coefficient for the $i$-th reflection at the $i$-th reflector on the path way.


\begin{figure}[t!]
  \centering
  \includegraphics[width=0.575\linewidth]{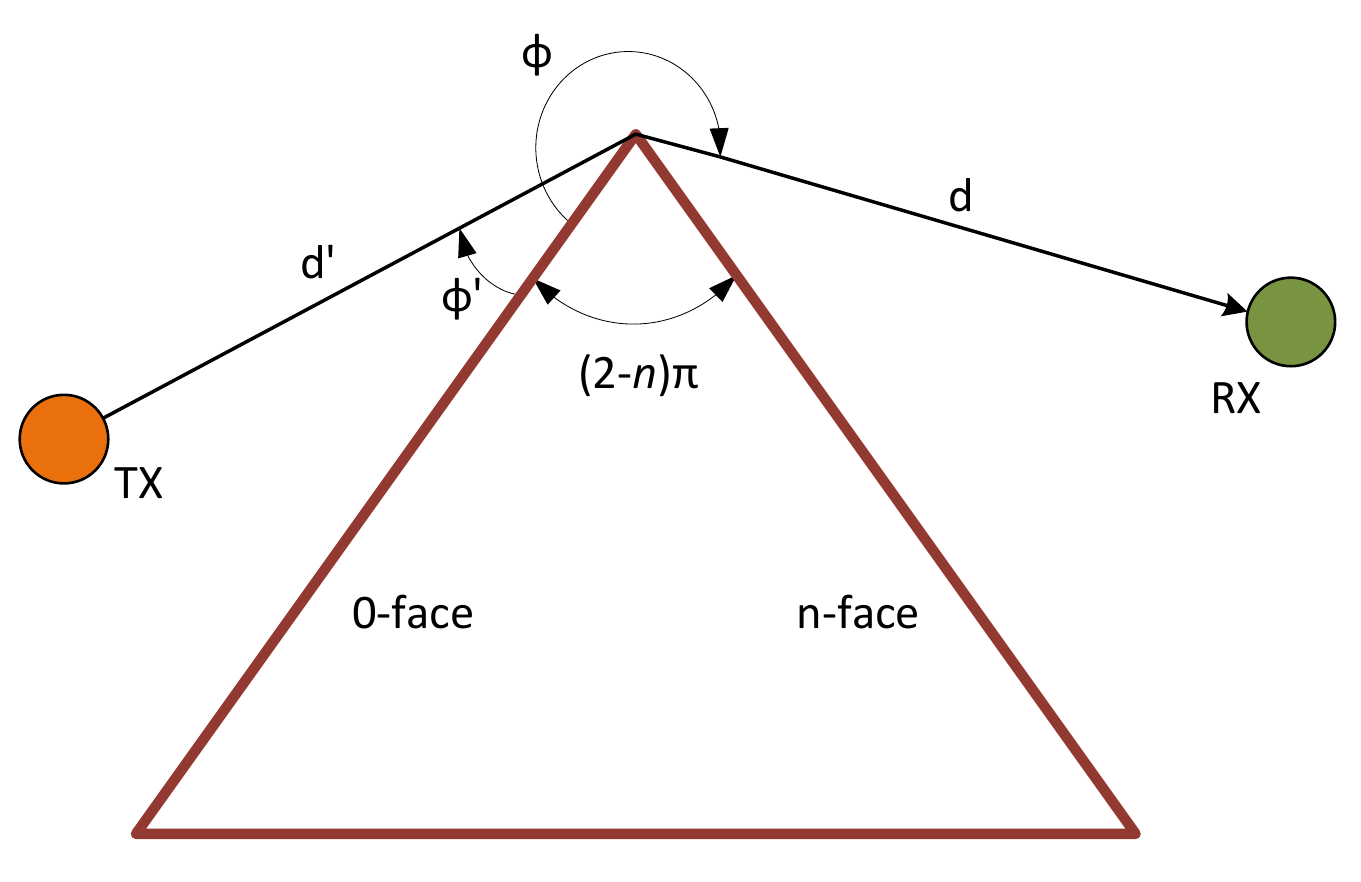}
  \vspace{-0.2cm}	
  \caption{Diffraction by a wedge}\label{fig:diffraction}
\end{figure}

\subsection{Scattering by Vegetation}

In the mmWave networks, especially in urban outdoor, vegetation will influence the propagation of mmWave signals, causing attenuation, scattering and depolarization.
According to radiative energy transfer~(RET) model~\cite{Leonor_TAP2014}, attenuation and scattering by vegetation, e.g. trees, can be described by a re-radiation function which specifies the amount of radiated energy transferred from one direction to another due to the random scattering phenomena.
The re-radiation pattern is angular dependent, and can be represented by a Gaussian-shaped forward lobe, superimposed to an isotropic background level representing the backscattering radiation.

Considering trees usually exhibit a cone like shape in nature, cylinders with varying sizes are adopted to represent trees as in~\cite{Leonor_TAP2014}.
The re-radiation pattern is given by
\begin{align}
\rho(\varphi) = {\alpha \left( {2}/{\beta}\right)^2 e^{-\left({\varphi}/{\beta}\right)^2}+ (1-\alpha)},
\end{align}
where $\varphi$~(in radians) is the scattering direction related to the incident direction as shown in Fig.~\ref{fig:scattering}, $\beta$ represents the 3-dB width of the forward lobe and $\alpha$ is the ratio of the forward scattered to the total scatter power.
The power density at the scatterer is given by
$S ={P_{\rm t}}/{4\pi r_1^2}$,
with $P_{\rm t}$ the transmit power and  so the power at the receiver is then given by
\begin{align}
P_{\rm r} = (1-\chi) 2rh S \rho(\varphi) \left( \frac{\lambda}{4\pi r_2}\right)^2,
\end{align}
where $\chi$ is the portion of energy absorbed by the tree, $r$ and $h$ the radius and height of the tree cylinder, $r_1$ is the distance between TX and scatter, $r_2$ is the distance between RX and scatter as depicted in Fig.~\ref{fig:scattering}.
The received power is assumed to be contributed by $n_{\rm S}$ uncorrelated rays with random phases and the path coefficient for a scattered ray can be modeled as
\begin{align}
\alpha_{\rm Scat} = e^{j\zeta}\sqrt{\frac{1}{n_{\rm S}}\cdot\frac{P_{\rm r}}{P_{\rm t}}} =  e^{j\zeta}\sqrt{\frac{(1-\chi)rh \rho(\varphi)}{32n_{\rm S}\pi^3}} \frac{\lambda}{r_1 r_2},
\end{align}
where $e^{j\zeta}$ is a random phase. In OmniSIM, only the first order scattering is considered as the scattering power by vegetation is limited compared with LoS and reflections by buildings.
Finally, the intra-cluster parameters of delays, DoAs and DoDs are generated in a similar way as in Section~\ref{sec:ref}.

\subsection{Diffraction Rays by Wedges}

Although diffraction is limited in mmWave frequencies, it should be considered for obstructed-line-of-sight~(OLOS) links where RX is close to the shadowing boundaries
by building wedges to accurately model the transition from LoS to NLOS condition.
According to UTD~\cite{Holm2000}, the diffracted rays are multiplied by diffraction coefficients and the diffraction coefficient by a wedge is given by
\begin{align}\label{eq1}
D = D(L,n;\phi,\phi^{\prime}) = D_1 + D_2 + \Gamma_0 D_{3} + \Gamma_n D_{4},
\end{align}
where $\Gamma_0$ and $\Gamma_n$ are reflection coefficients for the zero- and $n$-face, respectively, as shown in Fig.~\ref{fig:diffraction}.
The components $D_{l} (l=1,2,3,4)$ of the
diffraction coefficient are given by
\begin{equation}\label{cc}
\begin{aligned}
D_{l}(L,n;\phi,\phi^{\prime})\!=\!\frac{-e^{-j\pi/4}}{2n\sqrt{2\pi k}} \cot \gamma_{l} F(2kLn^2\!\sin^2\!\gamma_{l}),
\end{aligned}
\end{equation}
where $k$ is the wave number, $L = d^\prime d /(d^\prime + d)$, $n\pi$ is the exterior angle of the wedge, $F(\cdot)$ is the transition function defined in \cite{Kouyoumjian1974}, and $\gamma_{1} = [\pi - (\phi - \phi^{\prime})]/2n$, $\gamma_{2} = [\pi + (\phi - \phi^{\prime})]/2n$, $\gamma_{3} = [\pi - (\phi + \phi^{\prime})]/2n$, $\gamma_{4} = [\pi + (\phi + \phi^{\prime})]/2n$.
The basic idea of the UTD is that diffracted rays can be treated
in the same way as reflected rays in the GO and the path coefficient for the diffracted ray in Fig.~\ref{fig:diffraction} can be written as
\begin{equation}\label{ww}
\begin{aligned}
\alpha_{\rm Diff}
= \frac{e^{-jkd\prime}}{d^\prime} D \sqrt{\frac{d^\prime}{dd_{\rm T}}}e^{-jkd} = \frac{e^{-jkd_{\rm T}}}{d_{\rm T}} D \sqrt{\frac{d_{\rm T}}{d^\prime d}},
\end{aligned}
\end{equation}
where $d_{\rm T} = d + d^\prime$ is the total path distance.
In OmniSIM, only the first order diffraction is considered as the diffraction power is limited compared with LoS and reflections by buildings, and the diffraction is
characterized by a single ray.

\begin{table}[t!]
\centering
\caption{Simulation Parameters.}
\label{tbl:tab}	
\begin{tabular}{@{}ll|ll@{}}
\toprule
{UE velocity $v$} & 2~m/s & Sub-ray number $n_S$  & 20 \\
{Relative permittivity $\varepsilon$} & 2\,-\,6   & Intra-cluster DS  $\rm D_S$  & 12\,ns  \\
Surface roughtness $\sigma_{\rm h}$   & 0.4\m &    Azimuth  AS        & $10^\circ$ \\
Reradiation beamwidth  $\beta$       & $20^\circ$ &    Elevation AS    & $5^\circ$ \\
Scattered power ratio   $\alpha$          & 0.5 &    Absorption factor $\chi$     & 0.6\\
Polarization           & Vertical                     &  Direct ray spacing       &  $0.1^\circ$ \\
Radius of a tree $r$        & 4\m               &  Height of a tree $h$           &  5\m \\
BS height $h_{\rm BS}$      & 8\m          &  UE height        $h_{\rm UE}$  &  1.5\m \\
BS antenna element           & Microstrip          &  Carrier frequency  &  28\,GHz \\
RF bandwidth    & 246~MHz & TX Power & 30\,dBm \\
Subcarrier spacing & 120~kHz  &  Subcarrier number & 2048 \\
\bottomrule
\end{tabular}		
\vspace{-0.2cm}	
\end{table}

\section{Simulations}
In this section, we perform channel emulations via the developed OmniSIM in a dense urban scenario to show the capabilities of OmniSIM.
Fig.~\ref{fig:3Dmap} shows the bird-eye view of the simulated scenario. There are total 1606 building surfaces in the digital map.
We simulate the uplink channels from UE to BS. The BS is assumed to use an  uniform planar array ~(UPA), whose array response vector is given in~\cite{JiangWei2022}, with $16\times16$ microstrip patch antenna elements,  while an UE is equipped with an omni-directional antenna. The patch antenna radiation pattern of BS is shown in Fig.~\ref{fig:3Dmap}, which will be embedded in the BS array response vector. The pulse-shaping filter adopted in Eq.~\eqref{eq:ChannelGeneric} is $p(\tau) = {\rm{sinc}}\left({\pi \tau}/{T}\right)$. More details of the simulation parameters are given in Table~\ref{tbl:tab}. 
An UE trajectory is generated along the urban street which is shown in Fig.~\ref{fig:RTresult}. 

OmniSIM can generate wide-band channel responses for the 166 UE positions along the UE trajectory in 50~seconds using a 3.20\,GHz Intel i7-8700 CPU. The propagation paths, including reflection, diffraction and scattering paths for two UE positions $\rm A$ and $\rm B$, computed by the proposed ray-tracing method are depicted in Fig.~\ref{fig:RTresult}.  The corresponding joint angle-delay power profiles~(JADPP) observed at the BS from these two positions are shown in Fig.~\ref{fig:JADP}.
The power of the LoS ray $|\alpha_{\rm LoS}|^2$ at position 1 is 98.3~dB while that is 88.8 dB at position 2. For all UE positions in LoS condition to the BS, the power of LoS is at least 10~dB larger than other reflected, diffracted or scattered rays.
To verify the spatial consistency~\cite{NYU2018} of the channels generated by OmniSIM, the JADPPs for another two positions $\rm A^{\prime}$ and $\rm B^{\prime}$, following $\rm A$ and $\rm B$, are also shown in Fig.~\ref{fig:JADP}. It can be seen that position $\rm A^{\prime}$ exhibits a similar JADPP as $\rm A$, and $\rm B^{\prime}$ has a similar JADPP as $\rm B$.

The average channel power $\mathbb{E}_n\{|\mb{H}_{n,s}|^2\}$ over sub-carriers as a function of time is shown by Fig.~\ref{fig:ChPowers}. It can be seen that the channel power first increases as the UE moves towards the BS and then decreases after the UE moves into another street. At the end part of the UE trajectory, the mmWave channel is blocked by buildings and a sharp drop in channel power appears. This shows that OmniSIM can model the mmWave blockage effect inherently, which is important in mmWave cellular network planning.
\begin{figure}
  \centering
  \includegraphics[width=0.45\linewidth]{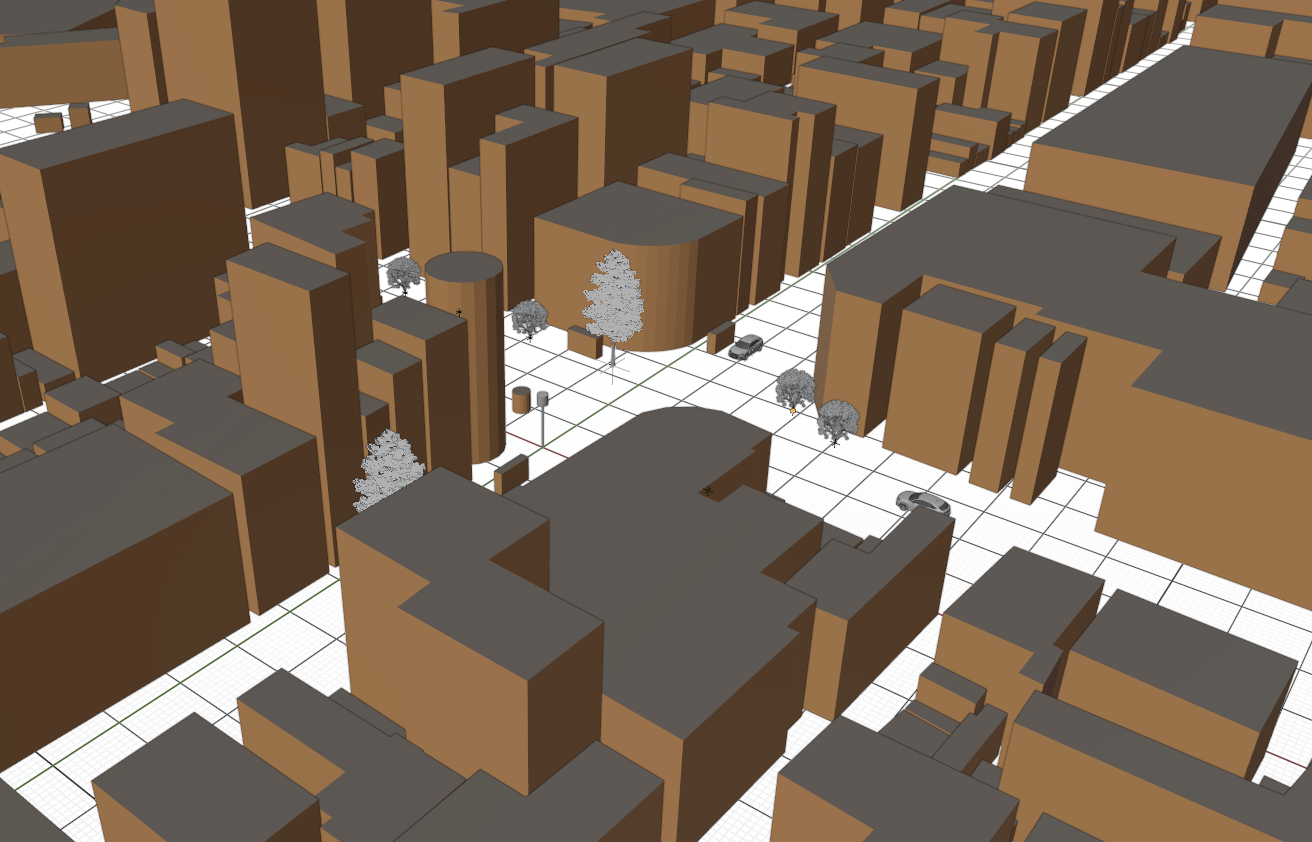}
  \includegraphics[width=0.45\linewidth]{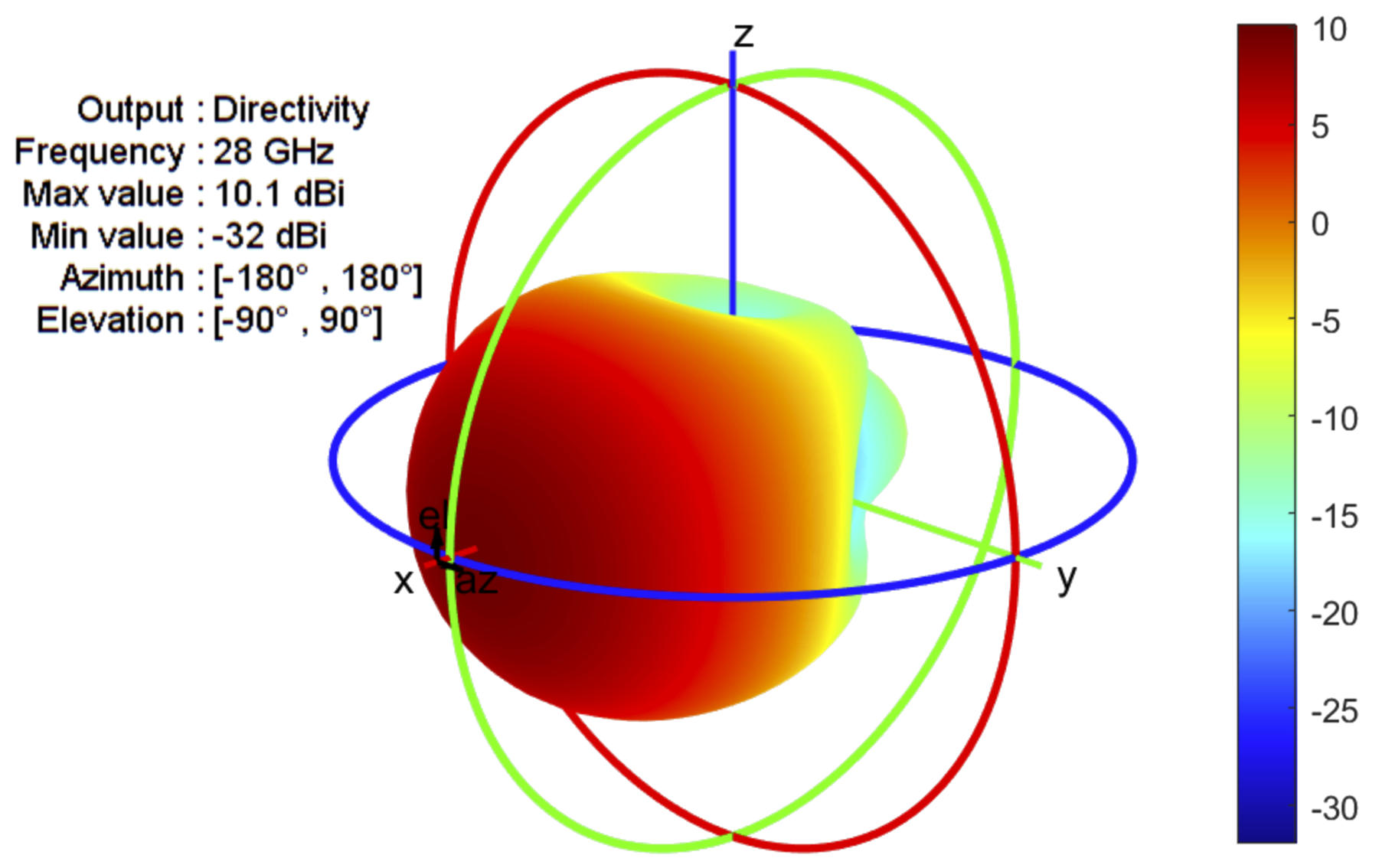}
  \vspace{-0.2cm}	
  \caption{3D view of the simulated scenario and BS antenna element pattern.}\label{fig:3Dmap}
\end{figure}

\begin{figure}
  \centering
  \includegraphics[width=0.75\linewidth]{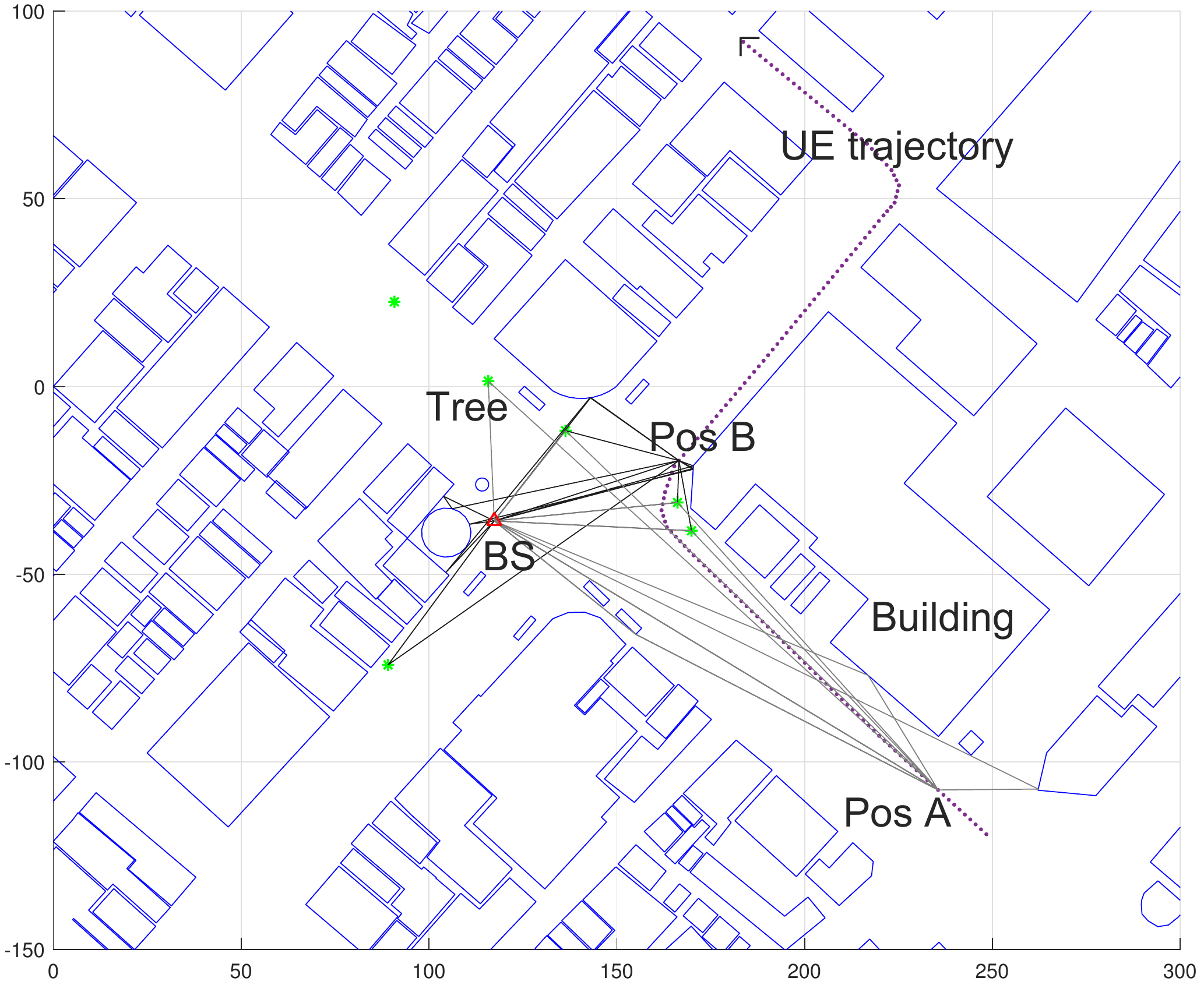}
  \vspace{-0.2cm}	
  \caption{Propagation Paths for two UE positions}\label{fig:RTresult}
\end{figure}

\begin{figure}
  \centering
  \includegraphics[width=0.4\linewidth]{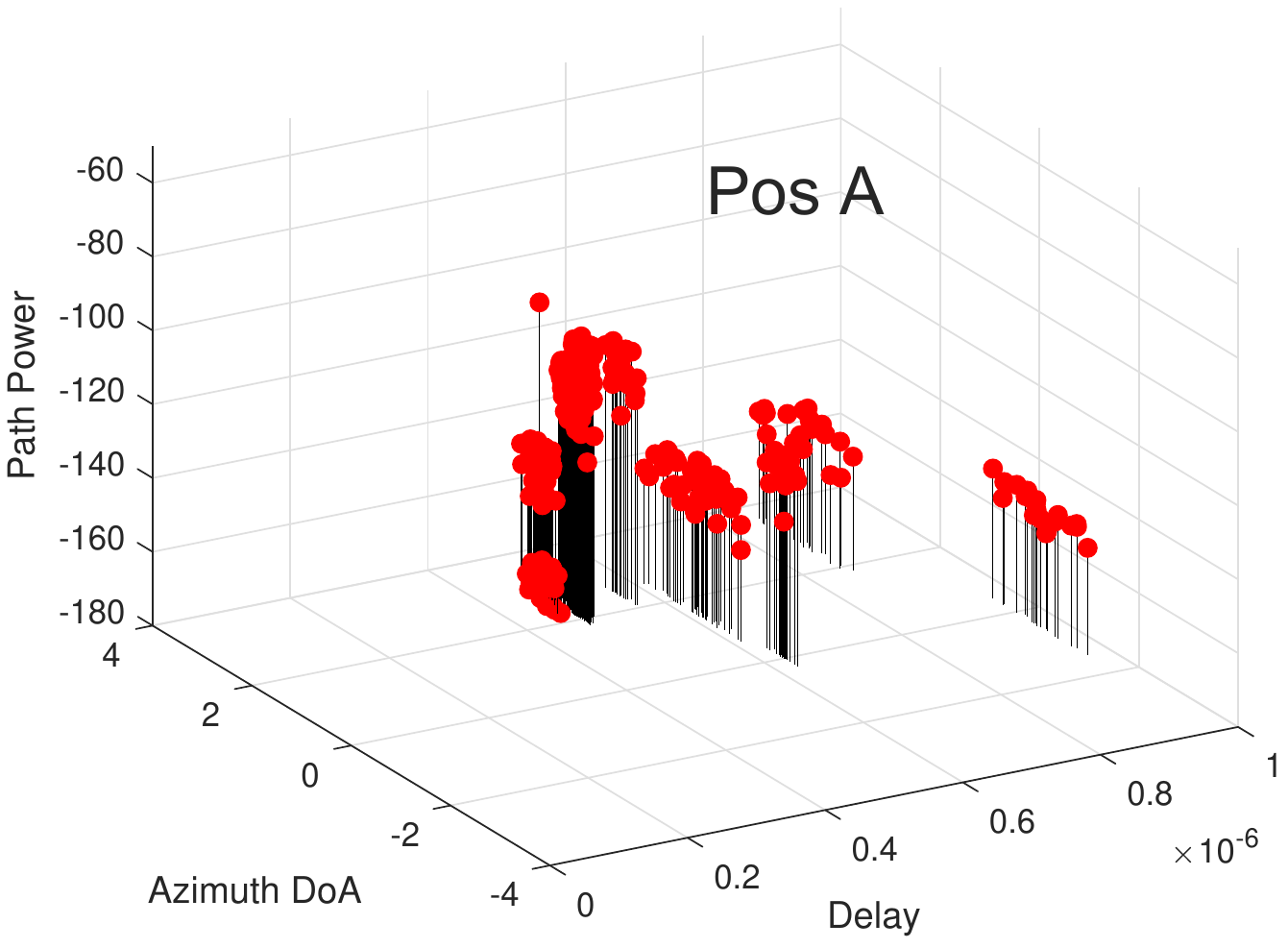}
  \includegraphics[width=0.4\linewidth]{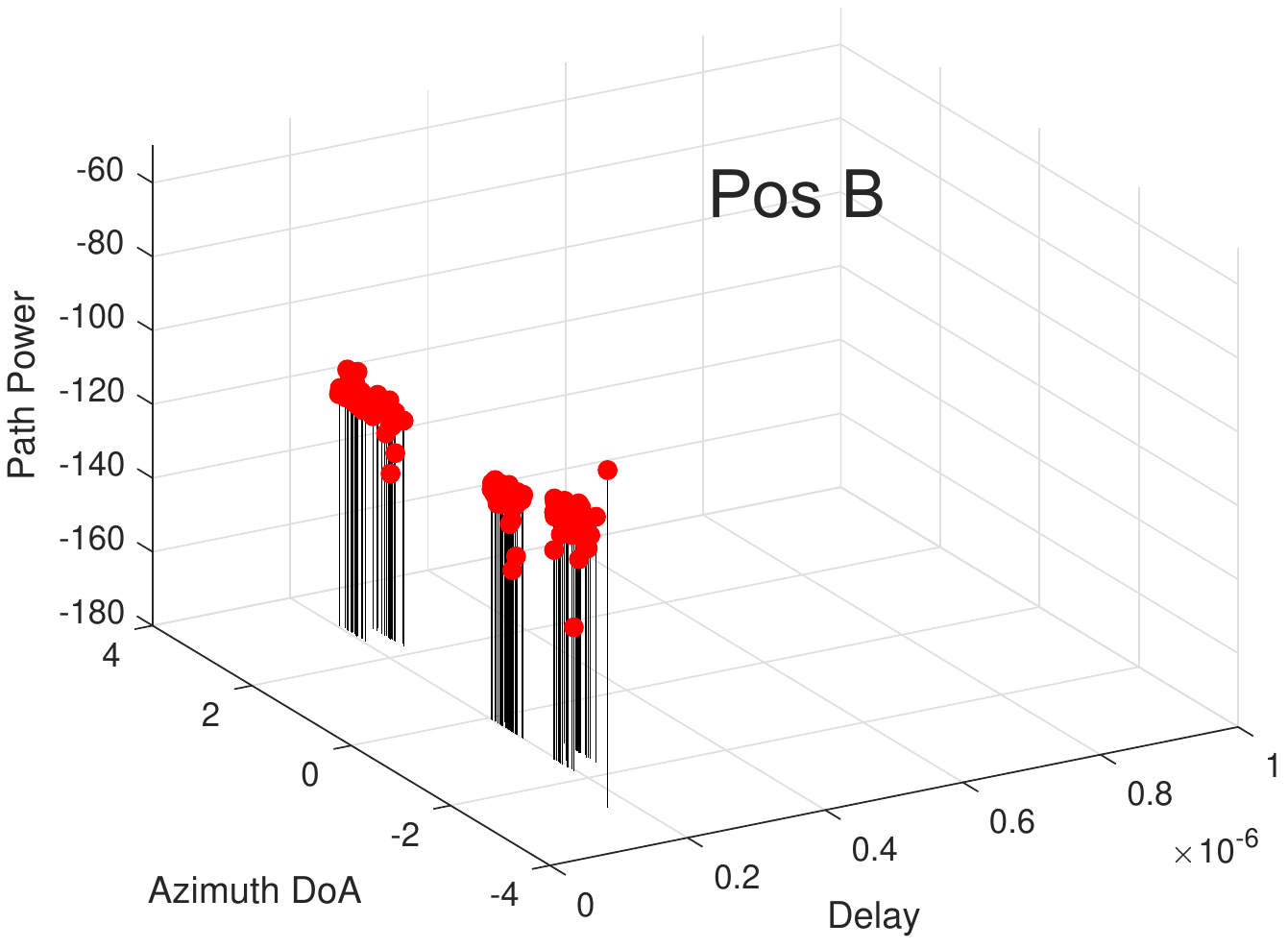}
  \includegraphics[width=0.4\linewidth]{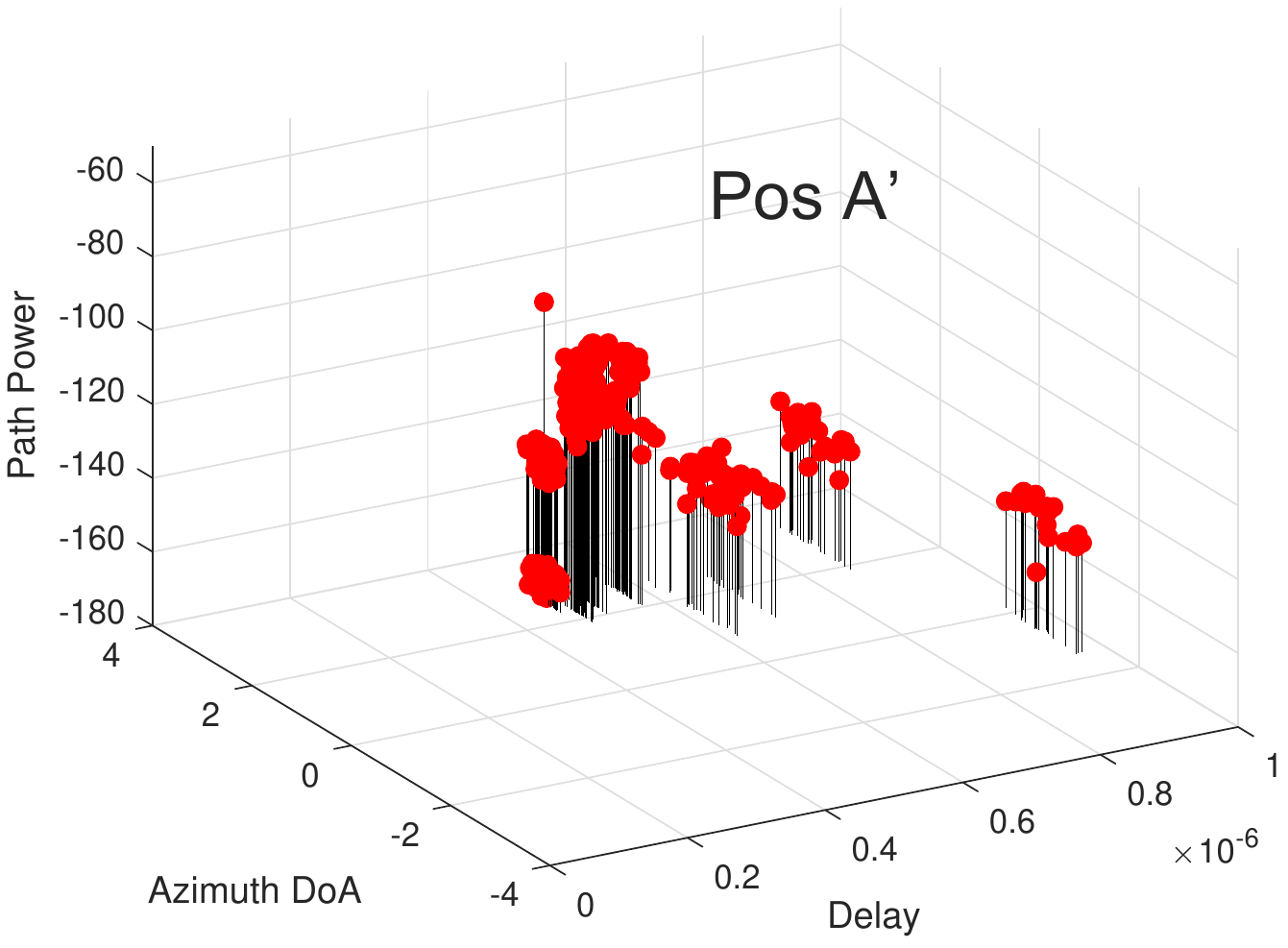}
  \includegraphics[width=0.4\linewidth]{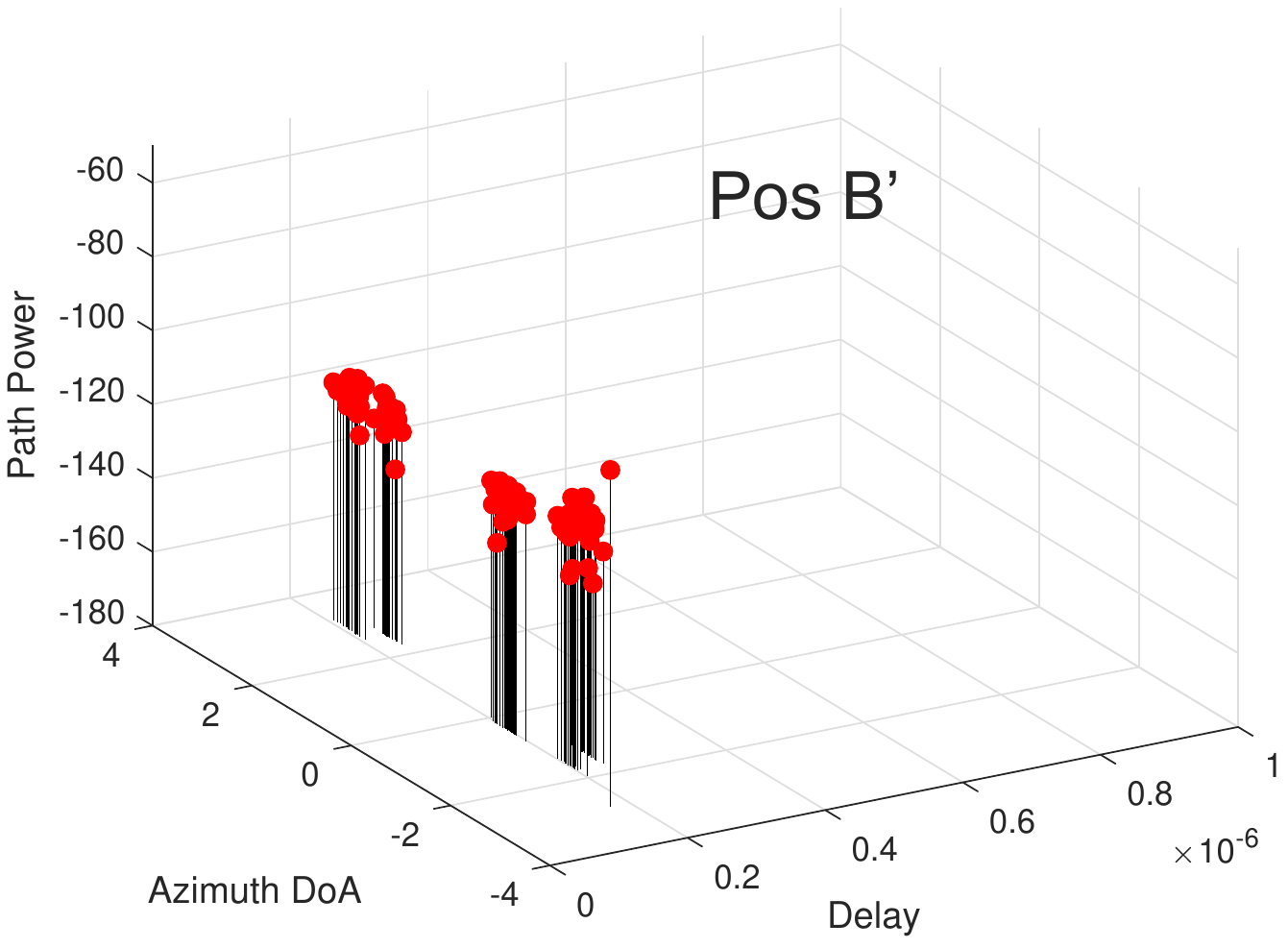}
  \vspace{-0.2cm}	
  \caption{Joint angle-delay power profile~(JADPP) for two UE positions}\label{fig:JADP}
\end{figure}

\begin{figure}
  \centering
  \includegraphics[width=0.75\linewidth]{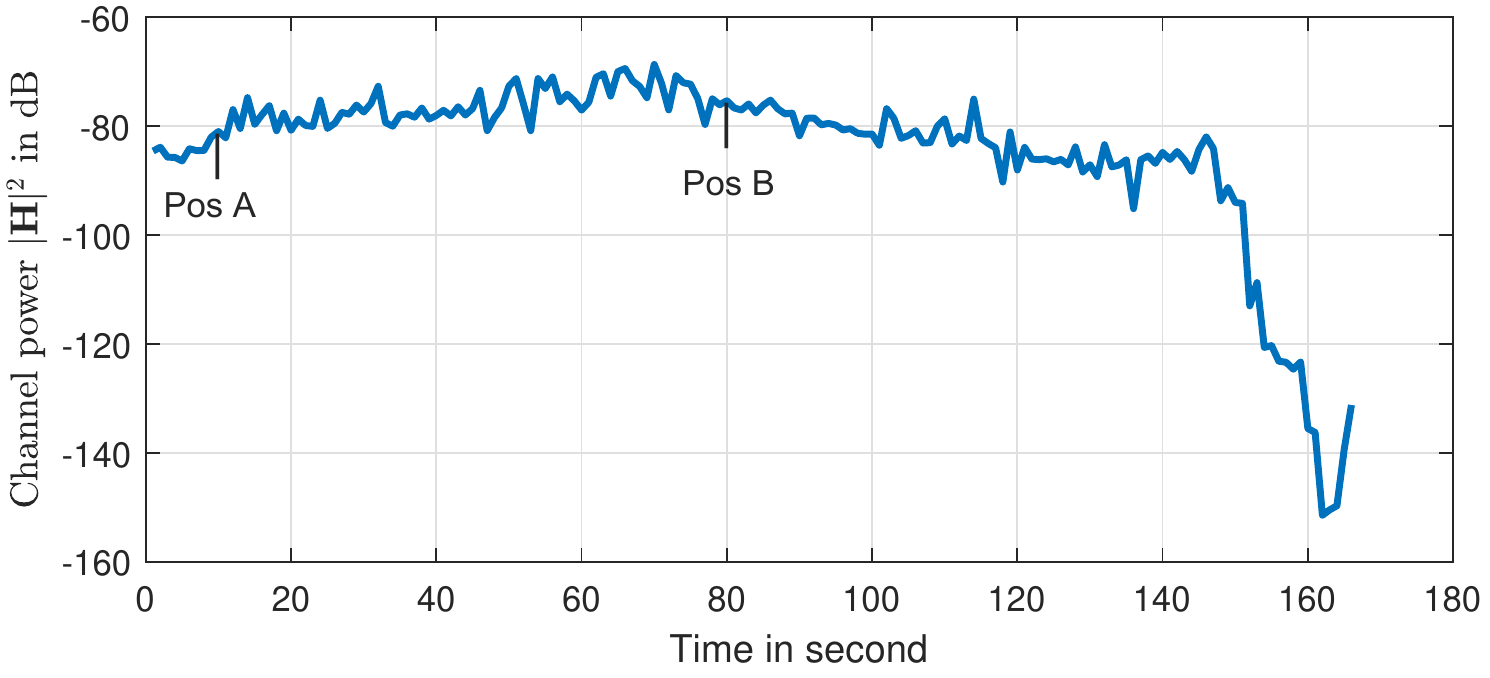}
  \vspace{-0.2cm}	
  \caption{Uplink channel powers along the UE trajectory}\label{fig:ChPowers}
\end{figure}

\section{Conclusions}
This paper has presented an open-source map-based hybrid millimeter-wave channel simulator OmniSIM which combines a customized ray tracing method with stochastic multi-path intra-cluster models widely adopted in GSCM. A fast shooting-bouncing rays algorithm has been developed to generate propagation paths for site-specific mmWave radio environments. Diffuse reflections by building surfaces, diffractions by wedges and scattering effects by vegetation have been properly modeled in OmniSIM. OmniSIM can generate spatial-consistent channels for network-level mmWave simulations with massive mmWave links. As the simulated channels are location-dependent, OmniSIM can be used for novel mmWave JCL research, such as intelligent beamforming and deep-learning based localization.

\bibliographystyle{IEEEtran}
\bibliography{iccc2022}

\end{document}

%% file: vmr-symbols-vecbold.tex
\usepackage{amssymb}
\usepackage{amsfonts}
\usepackage{mathrsfs}
\usepackage{xspace}
\usepackage{bm}
\usepackage{upgreek}

\newcommand{\safemath}[2]{\newcommand{#1}{\ensuremath{#2}\xspace}}

\safemath{\bma}{\mathbf{a}}
\safemath{\bmb}{\mathbf{b}}
\safemath{\bmc}{\mathbf{c}}
\safemath{\bmd}{\mathbf{d}}
\safemath{\bme}{\mathbf{e}}
\safemath{\bmf}{\mathbf{f}}
\safemath{\bmg}{\mathbf{g}}
\safemath{\bmh}{\mathbf{h}}
\safemath{\bmi}{\mathbf{i}}
\safemath{\bmj}{\mathbf{j}}
\safemath{\bmk}{\mathbf{k}}
\safemath{\bml}{\mathbf{l}}
\safemath{\bmm}{\mathbf{m}}
\safemath{\bmn}{\mathbf{n}}
\safemath{\bmo}{\mathbf{o}}
\safemath{\bmp}{\mathbf{p}}
\safemath{\bmq}{\mathbf{q}}
\safemath{\bmr}{\mathbf{r}}
\safemath{\bms}{\mathbf{s}}
\safemath{\bmt}{\mathbf{t}}
\safemath{\bmu}{\mathbf{u}}
\safemath{\bmv}{\mathbf{v}}
\safemath{\bmw}{\mathbf{w}}
\safemath{\bmx}{\mathbf{x}}
\safemath{\bmy}{\mathbf{y}}
\safemath{\bmz}{\mathbf{z}}
\safemath{\bmzero}{\mathbf{0}}
\safemath{\bmone}{\mathbf{1}}

\bmdefine{\biad}{a}
\bmdefine{\bibd}{b}
\bmdefine{\bicd}{c}
\bmdefine{\bidd}{d}
\bmdefine{\bied}{e}
\bmdefine{\bifd}{f}
\bmdefine{\bigd}{g}
\bmdefine{\bihd}{h}
\bmdefine{\biid}{i}
\bmdefine{\bijd}{j}
\bmdefine{\bikd}{k}
\bmdefine{\bild}{l}
\bmdefine{\bimd}{m}
\bmdefine{\bind}{n}
\bmdefine{\biod}{o}
\bmdefine{\bipd}{p}
\bmdefine{\biqd}{q}
\bmdefine{\bird}{r}
\bmdefine{\bisd}{s}
\bmdefine{\bitd}{t}
\bmdefine{\biud}{u}
\bmdefine{\bivd}{v}
\bmdefine{\biwd}{w}
\bmdefine{\bixd}{x}
\bmdefine{\biyd}{y}
\bmdefine{\bizd}{z}

\bmdefine{\bixid}{\xi}
\bmdefine{\bilambdad}{\lambda}
\bmdefine{\bimud}{\mu}
\bmdefine{\bithetad}{\theta}
\bmdefine{\biphid}{\phi}
\bmdefine{\bideltad}{\delta}

\safemath{\bmia}{\biad}
\safemath{\bmib}{\bibd}
\safemath{\bmic}{\bicd}
\safemath{\bmid}{\bidd}
\safemath{\bmie}{\bied}
\safemath{\bmif}{\bifd}
\safemath{\bmig}{\bigd}
\safemath{\bmih}{\bihd}
\safemath{\bmii}{\biid}
\safemath{\bmij}{\bijd}
\safemath{\bmik}{\bikd}
\safemath{\bmil}{\bild}
\safemath{\bmim}{\bimd}
\safemath{\bmin}{\bind}
\safemath{\bmio}{\biod}
\safemath{\bmip}{\bipd}
\safemath{\bmiq}{\biqd}
\safemath{\bmir}{\bird}
\safemath{\bmis}{\bisd}
\safemath{\bmit}{\bitd}
\safemath{\bmiu}{\biud}
\safemath{\bmiv}{\bivd}
\safemath{\bmiw}{\biwd}
\safemath{\bmix}{\bixd}
\safemath{\bmiy}{\biyd}
\safemath{\bmiz}{\bizd}

\safemath{\bmxi}{\bixid}
\safemath{\bmlambda}{\bilambdad}
\safemath{\bmmu}{\bimud}
\safemath{\bmtheta}{\bithetad}
\safemath{\bmphi}{\biphid}
\safemath{\bmdelta}{\bideltad}

\safemath{\bA}{\mathbf{A}}
\safemath{\bB}{\mathbf{B}}
\safemath{\bC}{\mathbf{C}}
\safemath{\bD}{\mathbf{D}}
\safemath{\bE}{\mathbf{E}}
\safemath{\bF}{\mathbf{F}}
\safemath{\bG}{\mathbf{G}}
\safemath{\bH}{\mathbf{H}}
\safemath{\bI}{\mathbf{I}}
\safemath{\bJ}{\mathbf{J}}
\safemath{\bK}{\mathbf{K}}
\safemath{\bL}{\mathbf{L}}
\safemath{\bM}{\mathbf{M}}
\safemath{\bN}{\mathbf{N}}
\safemath{\bO}{\mathbf{O}}
\safemath{\bP}{\mathbf{P}}
\safemath{\bQ}{\mathbf{Q}}
\safemath{\bR}{\mathbf{R}}
\safemath{\bS}{\mathbf{S}}
\safemath{\bT}{\mathbf{T}}
\safemath{\bU}{\mathbf{U}}
\safemath{\bV}{\mathbf{V}}
\safemath{\bW}{\mathbf{W}}
\safemath{\bX}{\mathbf{X}}
\safemath{\bY}{\mathbf{Y}}
\safemath{\bZ}{\mathbf{Z}}

\safemath{\bZero}{\mathbf{0}}
\safemath{\bOne}{\mathbf{1}}
\safemath{\bDelta}{\mathbf{\Delta}}
\safemath{\bLambda}{\mathbf{\UpLambda}}
\safemath{\bPhi}{\mathbf{\Upphi}}
\safemath{\bSigma}{\mathbf{\Upsigma}}
\safemath{\bOmega}{\mathbf{\Upomega}}
\safemath{\bTheta}{\mathbf{\Uptheta}}

\bmdefine{\biAd}{A}
\bmdefine{\biBd}{B}
\bmdefine{\biCd}{C}
\bmdefine{\biDd}{D}
\bmdefine{\biEd}{E}
\bmdefine{\biFd}{F}
\bmdefine{\biGd}{G}
\bmdefine{\biHd}{H}
\bmdefine{\biId}{I}
\bmdefine{\biJd}{J}
\bmdefine{\biKd}{K}
\bmdefine{\biLd}{L}
\bmdefine{\biMd}{M}
\bmdefine{\biOd}{N}
\bmdefine{\biPd}{O}
\bmdefine{\biQd}{P}
\bmdefine{\biRd}{R}
\bmdefine{\biSd}{S}
\bmdefine{\biTd}{T}
\bmdefine{\biUd}{U}
\bmdefine{\biVd}{V}
\bmdefine{\biWd}{W}
\bmdefine{\biXd}{X}
\bmdefine{\biYd}{Y}
\bmdefine{\biZd}{Z}

\bmdefine{\biDelta}{\Delta}
\bmdefine{\biLambda}{\Lambda}
\bmdefine{\biPhi}{\Phi}
\bmdefine{\biSigma}{\Sigma}
\bmdefine{\biOmega}{\Omega}
\bmdefine{\biTheta}{\Theta}

\safemath{\bimA}{\biAd}
\safemath{\bimB}{\biBd}
\safemath{\bimC}{\biCd}
\safemath{\bimD}{\biDd}
\safemath{\bimE}{\biEd}
\safemath{\bimF}{\biFd}
\safemath{\bimG}{\biGd}
\safemath{\bimH}{\biHd}
\safemath{\bimI}{\biId}
\safemath{\bimJ}{\biJd}
\safemath{\bimK}{\biKd}
\safemath{\bimL}{\biLd}
\safemath{\bimM}{\biMd}
\safemath{\bimN}{\biNd}
\safemath{\bimO}{\biOd}
\safemath{\bimP}{\biPd}
\safemath{\bimQ}{\biQd}
\safemath{\bimR}{\biRd}
\safemath{\bimS}{\biSd}
\safemath{\bimT}{\biTd}
\safemath{\bimU}{\biUd}
\safemath{\bimV}{\biVd}
\safemath{\bimW}{\biWd}
\safemath{\bimX}{\biXd}
\safemath{\bimY}{\biYd}
\safemath{\bimZ}{\biZd}

\safemath{\bimDelta}{\biDelta}
\safemath{\bimLambda}{\biLambda}
\safemath{\bimPhi}{\biPhi}
\safemath{\bimSigma}{\biSigma}
\safemath{\bimOmega}{\biOmega}
\safemath{\bimTheta}{\biTheta}

\safemath{\setA}{\mathcal{A}}
\safemath{\setB}{\mathcal{B}}
\safemath{\setC}{\mathcal{C}}
\safemath{\setD}{\mathcal{D}}
\safemath{\setE}{\mathcal{E}}
\safemath{\setF}{\mathcal{F}}
\safemath{\setG}{\mathcal{G}}
\safemath{\setH}{\mathcal{H}}
\safemath{\setI}{\mathcal{I}}
\safemath{\setJ}{\mathcal{J}}
\safemath{\setK}{\mathcal{K}}
\safemath{\setL}{\mathcal{L}}
\safemath{\setM}{\mathcal{M}}
\safemath{\setN}{\mathcal{N}}
\safemath{\setO}{\mathcal{O}}
\safemath{\setP}{\mathcal{P}}
\safemath{\setQ}{\mathcal{Q}}
\safemath{\setR}{\mathcal{R}}
\safemath{\setS}{\mathcal{S}}
\safemath{\setT}{\mathcal{T}}
\safemath{\setU}{\mathcal{U}}
\safemath{\setV}{\mathcal{V}}
\safemath{\setW}{\mathcal{W}}
\safemath{\setX}{\mathcal{X}}
\safemath{\setY}{\mathcal{Y}}
\safemath{\setZ}{\mathcal{Z}}
\safemath{\emptySet}{\varnothing}

\safemath{\colA}{\mathscr{A}}
\safemath{\colB}{\mathscr{B}}
\safemath{\colC}{\mathscr{C}}
\safemath{\colD}{\mathscr{D}}
\safemath{\colE}{\mathscr{E}}
\safemath{\colF}{\mathscr{F}}
\safemath{\colG}{\mathscr{G}}
\safemath{\colH}{\mathscr{H}}
\safemath{\colI}{\mathscr{I}}
\safemath{\colJ}{\mathscr{J}}
\safemath{\colK}{\mathscr{K}}
\safemath{\colL}{\mathscr{L}}
\safemath{\colM}{\mathscr{M}}
\safemath{\colN}{\mathscr{N}}
\safemath{\colO}{\mathscr{O}}
\safemath{\colP}{\mathscr{P}}
\safemath{\colQ}{\mathscr{Q}}
\safemath{\colR}{\mathscr{R}}
\safemath{\colS}{\mathscr{S}}
\safemath{\colT}{\mathscr{T}}
\safemath{\colU}{\mathscr{U}}
\safemath{\colV}{\mathscr{V}}
\safemath{\colW}{\mathscr{W}}
\safemath{\colX}{\mathscr{X}}
\safemath{\colY}{\mathscr{Y}}
\safemath{\colZ}{\mathscr{Z}}

\safemath{\opA}{\mathbb{A}}
\safemath{\opB}{\mathbb{B}}
\safemath{\opC}{\mathbb{C}}
\safemath{\opD}{\mathbb{D}}
\safemath{\opE}{\mathbb{E}}
\safemath{\opF}{\mathbb{F}}
\safemath{\opG}{\mathbb{G}}
\safemath{\opH}{\mathbb{H}}
\safemath{\opI}{\mathbb{I}}
\safemath{\opJ}{\mathbb{J}}
\safemath{\opK}{\mathbb{K}}
\safemath{\opL}{\mathbb{L}}
\safemath{\opM}{\mathbb{M}}
\safemath{\opN}{\mathbb{N}}
\safemath{\opO}{\mathbb{O}}
\safemath{\opP}{\mathbb{P}}
\safemath{\opQ}{\mathbb{Q}}
\safemath{\opR}{\mathbb{R}}
\safemath{\opS}{\mathbb{S}}
\safemath{\opT}{\mathbb{T}}
\safemath{\opU}{\mathbb{U}}
\safemath{\opV}{\mathbb{V}}
\safemath{\opW}{\mathbb{W}}
\safemath{\opX}{\mathbb{X}}
\safemath{\opY}{\mathbb{Y}}
\safemath{\opZ}{\mathbb{Z}}
\safemath{\opZero}{\mathbb{O}}
\safemath{\identityop}{\opI}

\safemath{\veca}{\bma}
\safemath{\vecb}{\bmb}
\safemath{\vecc}{\bmc}
\safemath{\vecd}{\bmd}
\safemath{\vece}{\bme}
\safemath{\vecf}{\bmf}
\safemath{\vecg}{\bmg}
\safemath{\vech}{\bmh}
\safemath{\veci}{\bmi}
\safemath{\vecj}{\bmj}
\safemath{\veck}{\bmk}
\safemath{\vecl}{\bml}
\safemath{\vecm}{\bmm}
\safemath{\vecn}{\bmn}
\safemath{\veco}{\bmo}
\safemath{\vecp}{\bmp}
\safemath{\vecq}{\bmq}
\safemath{\vecr}{\bmr}
\safemath{\vecs}{\bms}
\safemath{\vect}{\bmt}
\safemath{\vecu}{\bmu}
\safemath{\vecv}{\bmv}
\safemath{\vecw}{\bmw}
\safemath{\vecx}{\bmx}
\safemath{\vecy}{\bmy}
\safemath{\vecz}{\bmz}

\safemath{\veczero}{\bmzero}
\safemath{\vecone}{\bmone}
\safemath{\vecxi}{\bmxi}
\safemath{\veclambda}{\bmlambda}
\safemath{\vecmu}{\bmmu}
\safemath{\vectheta}{\bmtheta}
\safemath{\vecphi}{\bmphi}
\safemath{\vecdelta}{\bmdelta}

\safemath{\matA}{\bA}
\safemath{\matB}{\bB}
\safemath{\matC}{\bC}
\safemath{\matD}{\bD}
\safemath{\matE}{\bE}
\safemath{\matF}{\bF}
\safemath{\matG}{\bG}
\safemath{\matH}{\bH}
\safemath{\matI}{\bI}
\safemath{\matJ}{\bJ}
\safemath{\matK}{\bK}
\safemath{\matL}{\bL}
\safemath{\matM}{\bM}
\safemath{\matN}{\bN}
\safemath{\matO}{\bO}
\safemath{\matP}{\bP}
\safemath{\matQ}{\bQ}
\safemath{\matR}{\bR}
\safemath{\matS}{\bS}
\safemath{\matT}{\bT}
\safemath{\matU}{\bU}
\safemath{\matV}{\bV}
\safemath{\matW}{\bW}
\safemath{\matX}{\bX}
\safemath{\matY}{\bY}
\safemath{\matZ}{\bZ}
\safemath{\matzero}{\bmzero}

\safemath{\matDelta}{\bDelta}
\safemath{\matLambda}{\bLambda}
\safemath{\matPhi}{\bPhi}
\safemath{\matSigma}{\bSigma}
\safemath{\matOmega}{\bOmega}
\safemath{\matTheta}{\bTheta}

\safemath{\matidentity}{\matI}
\safemath{\matone}{\matO}

\safemath{\rnda}{A}
\safemath{\rndb}{B}
\safemath{\rndc}{C}
\safemath{\rndd}{D}
\safemath{\rnde}{E}
\safemath{\rndf}{F}
\safemath{\rndg}{G}
\safemath{\rndh}{H}
\safemath{\rndi}{I}
\safemath{\rndj}{J}
\safemath{\rndk}{K}
\safemath{\rndl}{L}
\safemath{\rndm}{M}
\safemath{\rndn}{N}
\safemath{\rndo}{O}
\safemath{\rndp}{P}
\safemath{\rndq}{Q}
\safemath{\rndr}{R}
\safemath{\rnds}{S}
\safemath{\rndt}{T}
\safemath{\rndu}{U}
\safemath{\rndv}{V}
\safemath{\rndw}{W}
\safemath{\rndx}{X}
\safemath{\rndy}{Y}
\safemath{\rndz}{Z}

\safemath{\rveca}{\bimA}
\safemath{\rvecb}{\bimB}
\safemath{\rvecc}{\bimC}
\safemath{\rvecd}{\bimD}
\safemath{\rvece}{\bimE}
\safemath{\rvecf}{\bimF}
\safemath{\rvecg}{\bimG}
\safemath{\rvech}{\bimH}
\safemath{\rveci}{\bimI}
\safemath{\rvecj}{\bimJ}
\safemath{\rveck}{\bimK}
\safemath{\rvecl}{\bimL}
\safemath{\rvecm}{\bimM}
\safemath{\rvecn}{\bimN}
\safemath{\rveco}{\bomO}
\safemath{\rvecp}{\bimP}
\safemath{\rvecq}{\bimQ}
\safemath{\rvecr}{\bimR}
\safemath{\rvecs}{\bimS}
\safemath{\rvect}{\bimT}
\safemath{\rvecu}{\bimU}
\safemath{\rvecv}{\bimV}
\safemath{\rvecw}{\bimW}
\safemath{\rvecx}{\bimX}
\safemath{\rvecy}{\bimY}
\safemath{\rvecz}{\bimZ}

\safemath{\rvecxi}{\bmxi}
\safemath{\rveclambda}{\bmlambda}
\safemath{\rvecmu}{\bmmu}
\safemath{\rvectheta}{\bmtheta}
\safemath{\rvecphi}{\bmphi}

\safemath{\rmatA}{\bimA}
\safemath{\rmatB}{\bimB}
\safemath{\rmatC}{\bimC}
\safemath{\rmatD}{\bimD}
\safemath{\rmatE}{\bimE}
\safemath{\rmatF}{\bimF}
\safemath{\rmatG}{\bimG}
\safemath{\rmatH}{\bimH}
\safemath{\rmatI}{\bimI}
\safemath{\rmatJ}{\bimJ}
\safemath{\rmatK}{\bimK}
\safemath{\rmatL}{\bimL}
\safemath{\rmatM}{\bimM}
\safemath{\rmatN}{\bimN}
\safemath{\rmatO}{\bimO}
\safemath{\rmatP}{\bimP}
\safemath{\rmatQ}{\bimQ}
\safemath{\rmatR}{\bimR}
\safemath{\rmatS}{\bimS}
\safemath{\rmatT}{\bimT}
\safemath{\rmatU}{\bimU}
\safemath{\rmatV}{\bimV}
\safemath{\rmatW}{\bimW}
\safemath{\rmatX}{\bimX}
\safemath{\rmatY}{\bimY}
\safemath{\rmatZ}{\bimZ}

\safemath{\rmatDelta}{\bimDelta}
\safemath{\rmatLambda}{\bimLambda}
\safemath{\rmatPhi}{\bimPhi}
\safemath{\rmatSigma}{\bimSigma}
\safemath{\rmatOmega}{\bimOmega}
\safemath{\rmatTheta}{\bimTheta}

%% file: standard-macros.tex
\usepackage{amssymb}
\usepackage{amsfonts}
\usepackage{mathrsfs}
\usepackage{xspace}
\usepackage{bm}
\usepackage{fancyref}
\usepackage{textcomp}

\usepackage{multirow}
\usepackage{stmaryrd}

\newenvironment{textbmatrix}{	\setlength{\arraycolsep}{2.5pt}%
								\big[\begin{matrix}}{\end{matrix}\big]%
								\raisebox{0.08ex}{\vphantom{M}}}

\def\be{\begin{equation}}
\def\ee{\end{equation}}
\def\een{\nonumber \end{equation}}
\def\mat{\begin{bmatrix}}
\def\emat{\end{bmatrix}}
\def\btm{\begin{textbmatrix}}
\def\etm{\end{textbmatrix}}

\def\ba#1\ea{\begin{align}#1\end{align}}
\def\bas#1\eas{\begin{align*}#1\end{align*}}
\def\bs#1\es{\begin{split}#1\end{split}} 
\def\bg#1\eg{\begin{gather}#1\end{gather}}
\def\bml#1\eml{\begin{multline}#1\end{multline}}
\def\bi#1\ei{\begin{itemize}#1\end{itemize}}

\safemath{\dirac}{\delta}					
\safemath{\krond}{\dirac}					

\safemath{\upto}{\uparrow}
\safemath{\downto}{\downarrow}
\safemath{\iu}{j}							
\safemath{\ev}{\lambda}						
\safemath{\hilseqspace}{l^{2}}				
\newcommand{\banachfunspace}[1]{\setL^{#1}}	
\safemath{\hilfunspace}{\banachfunspace{2}}	

\safemath{\SNR}{\textsf{SNR}} 				
\safemath{\PAR}{\textsf{PAR}} 				
\safemath{\No}{N_0}							
\safemath{\Es}{E_s}							
\safemath{\Eb}{E_b}							
\safemath{\EbNo}{\frac{\Eb}{\No}}
\safemath{\EsNo}{\frac{\Es}{\No}}

\DeclareMathOperator{\CHop}{\ensuremath{\opH}} 
\safemath{\tvir}{\rndh_{\CHop}}				
\safemath{\tvtf}{\rndl_{\CHop}}				
\safemath{\spf}{\rnds_{\CHop}}				
\safemath{\bff}{H_{\CHop}}					

\safemath{\ircf}{r_{h}}						
\safemath{\tftvcf}{r_{s}}					
\safemath{\tfcf}{r_{l}}						
\safemath{\bfcf}{r_{H}}						

\safemath{\tcorr}{c_h}						
\safemath{\scf}{c_{s}}						
\safemath{\tfcorr}{c_{l}}					
\safemath{\fcorr}{c_{H}}						

\safemath{\mi}{I}							
\safemath{\capacity}{C}						

\safemath{\normal}{\mathcal{N}}			
\safemath{\jpg}{\mathcal{CN}}			
\safemath{\mchain}{\leftrightarrow}		

\safemath{\dB}{\,\mathrm{dB}}
\safemath{\dBm}{\,\mathrm{dBm}}
\safemath{\Hz}{\,\mathrm{Hz}}
\safemath{\kHz}{\,\mathrm{kHz}}
\safemath{\MHz}{\,\mathrm{MHz}}
\safemath{\GHz}{\,\mathrm{GHz}}
\safemath{\s}{\,\mathrm{s}}
\safemath{\ms}{\,\mathrm{ms}}
\safemath{\mus}{\,\mathrm{\text{\textmu}s}}
\safemath{\ns}{\,\mathrm{ns}}
\safemath{\ps}{\,\mathrm{ps}}
\safemath{\meter}{\,\mathrm{m}}
\safemath{\mm}{\,\mathrm{mm}}
\safemath{\cm}{\,\mathrm{cm}}
\safemath{\m}{\,\mathrm{m}}
\safemath{\W}{\,\mathrm{W}}
\safemath{\mW}{\, \mathrm{mW}}
\safemath{\J}{\,\mathrm{J}}
\safemath{\K}{\,\mathrm{K}}
\safemath{\bit}{\,\mathrm{bit}}
\safemath{\nat}{\,\mathrm{nat}}

\safemath{\define}{\triangleq}			

\safemath{\equivalent}{\sim}
\safemath{\distas}{\sim}					
\safemath{\sdiff}{\Delta}				

\safemath{\reals}{\mathbb{R}}
\safemath{\positivereals}{\reals_{+}}
\safemath{\integers}{\mathbb{Z}}
\safemath{\posint}{\integers_{+}}
\safemath{\naturals}{\mathbb{N}}
\safemath{\posnaturals}{\naturals_{+}}
\safemath{\complexset}{\mathbb{C}}
\safemath{\rationals}{\mathbb{Q}}

\newcommand*{\fancyrefapplabelprefix}{app}		
\newcommand*{\fancyrefthmlabelprefix}{thm}		
\newcommand*{\fancyreflemlabelprefix}{lem}		
\newcommand*{\fancyrefcorlabelprefix}{cor}		
\newcommand*{\fancyrefdeflabelprefix}{def}		
\newcommand*{\fancyrefproplabelprefix}{prop}	
\newcommand*{\fancyrefobslabelprefix}{obs}		
\newcommand*{\fancyrefalglabelprefix}{alg}		
\newcommand*{\fancyrefasmlabelprefix}{asm}	    
\newcommand*{\fancyrefasmslabelprefix}{asms}	    
\newcommand*{\fancyreftbllabelprefix}{tbl}	    
\newcommand*{\fancyreftremabelprefix}{rem}	    

\frefformat{vario}{\fancyrefseclabelprefix}{Section~#1}
\frefformat{vario}{\fancyrefthmlabelprefix}{Theorem~#1}
\frefformat{vario}{\fancyreflemlabelprefix}{Lemma~#1}
\frefformat{vario}{\fancyrefcorlabelprefix}{Corollary~#1}
\frefformat{vario}{\fancyrefdeflabelprefix}{Definition~#1}
\frefformat{vario}{\fancyrefobslabelprefix}{Observation~#1}
\frefformat{vario}{\fancyrefasmlabelprefix}{Assumption~#1}
\frefformat{vario}{\fancyrefasmslabelprefix}{Assumptions~#1}
\frefformat{vario}{\fancyreffiglabelprefix}{Figure~#1}
\frefformat{vario}{\fancyrefapplabelprefix}{Appendix~#1} 
\frefformat{vario}{\fancyrefproplabelprefix}{Proposition~#1}
\frefformat{vario}{\fancyrefalglabelprefix}{Algorithm~#1}
\frefformat{vario}{\fancyrefeqlabelprefix}{(#1)}
\frefformat{vario}{\fancyreftbllabelprefix}{Table~#1}
\frefformat{vario}{\fancyreftremabelprefix}{Remark~#1}

%% file: defs.tex
\safemath{\dictab}{[\,\dicta\,\,\dictb\,]}

\safemath{\ysig}{\bmy}
\safemath{\ysighat}{\hat{\ysig}}
\safemath{\ysigdim}{M}
\safemath{\xsig}{\bmx}
\safemath{\xsigdim}{N}
\safemath{\nx}{n_x}
\safemath{\zsig}{\bmz}
\safemath{\zsigdim}{\ysigdim}
\safemath{\rsig}{\bmr}
\safemath{\Adict}{\bA}
\safemath{\Adicttilde}{\widetilde{\Adict}}
\safemath{\Adictdim}{\outputdim\times\xsigdim}
\safemath{\avec}{\bma}
\safemath{\avectilde}{\tilde{\avec}}
\safemath{\Bdict}{\bB}
\safemath{\Bdicttilde}{\widetilde{\Bdict}}
\safemath{\Cdict}{\bC}
\safemath{\cvec}{\bmc}
\safemath{\Ddict}{\bD}
\safemath{\Ddictdim}{\ysigdim\times\xsigdim}
\safemath{\dvec}{\bmd}
\safemath{\Ddicttilde}{\widetilde{\bD}}
\safemath{\Bonb}{\bB}
\safemath{\bvec}{\bmb}
\safemath{\Bonbdim}{\ysigdim\times\ysigdim}
\safemath{\noise}{\bmn}
\safemath{\noisedim}{\ysigim}
\safemath{\err}{\bme}
\safemath{\errdim}{\ysigdim}
\safemath{\errset}{\setE}
\safemath{\nerr}{n_e}
\safemath{\delop}{\bP_\errset}
\safemath{\delopc}{\bP_{{\errset}^c}}

\safemath{\cplxi}{\imath}
\safemath{\cplxj}{\jmath}

\safemath{\dict}{\matD}
\safemath{\inputdim}{N}		
\safemath{\outputdim}{M}		
\safemath{\sparsity}{S}	
\safemath{\inputdimA}{{N_a}}	
\safemath{\inputdimB}{{N_b}}	
\safemath{\elemA}{{n_a}}	
\safemath{\elemB}{{n_b}}	
\safemath{\resA}{\matR_a}	
\safemath{\resB}{\matR_b}	
\safemath{\subD}{\matS} 
\safemath{\subA}{\matS_a} 
\safemath{\subB}{\matS_b} 
\safemath{\dicta}{\matA} 	
\safemath{\dictb}{\matB} 	
\safemath{\hollowS}{H}
\safemath{\hollowA}{H_a}
\safemath{\hollowB}{H_b}
\safemath{\cross}{Z}
\safemath{\coh}{\mu_d}			
\safemath{\coha}{\mu_a}			
\safemath{\cohb}{\mu_b}			
\safemath{\mubs}{\nu}	
\safemath{\cohm}{\mu_m} 
\safemath{\dictset}{\setD}	
\safemath{\dictsetp}{\dictset(\coh,\coha,\cohb)}	
\safemath{\dictsetgen}{\dictset_\text{gen}}
\safemath{\dictsetgenp}{\dictsetgen(\coh)}
\safemath{\dictsetonb}{\dictset_\text{onb}}
\safemath{\dictsetonbp}{\dictsetonb(\coh)}

\safemath{\leftside}{U}
\safemath{\rightsideA}{R_a}
\safemath{\rightsideB}{R_b}

\safemath{\indexS}{\setI_S} 

\safemath{\na}{n_a}			
\safemath{\nb}{n_b}			
\safemath{\coeffa}{p_i}	
\safemath{\coeffb}{q_j}	
\safemath{\seta}{\setP}		
\safemath{\setb}{\setQ}     
\safemath{\setw}{\setW}	
\safemath{\setz}{\setZ}	
\safemath{\cola}{\veca}		
\safemath{\colb}{\vecb}		
\safemath{\cold}{\vecd}		
\safemath{\inputvec}{\vecx} 	
\safemath{\error}{\vece}	
\safemath{\noiseout}{\vecz} 	
\safemath{\inputvecel}{x}
\safemath{\inputveca}{\vecx_a}
\safemath{\inputvecb}{\vecx_b}
\safemath{\outputvec}{\vecy}	
\safemath{\lambdamin}{\lambda_{\mathrm{min}}}

\safemath{\elltwo}{\ell_2}
\safemath{\ellone}{\ell_1}
\safemath{\ellzero}{\ell_0}
\safemath{\ellinf}{\ell_\infty}
\safemath{\ellinftilde}{\ell_{\widetilde\infty}}
\safemath{\licard}{Z(\coh,\coha,\cohb)}
\safemath{\xsol}{\hat{x}}
\safemath{\xbord}{x_b}		
\safemath{\xstat}{x_s}		
\safemath{\xstatLone}{\tilde{x}_s}
\safemath{\order}{\mathcal{O}} 
\safemath{\scales}{\Theta} 
\safemath{\ones}{\mathbf{1}} 
\safemath{\zeroes}{\mathbf{0}} 
\safemath{\thlone}{\kappa(\coh,\cohb)} 
\safemath{\constoneA}{\delta} 
\safemath{\constoneB}{\epsilon} 
\safemath{\nlarge}{L}				   
\safemath{\sumlarge}{S_\nlarge}
\safemath{\maxlarger}{P_\nlarge}	   
\safemath{\Pzero}{\textrm{P0}}	
\safemath{\Pone}{\textrm{P1}}
\safemath{\vecfir}{\vecw}			 
\safemath{\vecsec}{\vecz}
\safemath{\elvecfir}{w}              
\safemath{\elvecsec}{z}				 
\safemath{\nlargefir}{n}
\safemath{\normout}{\gamma}
\safemath{\auxfun}{h}
\safemath{\supp}{\textrm{supp}}

\safemath{\indexa}{\ell}
\safemath{\indexb}{r}
\safemath{\indexc}{i}
\safemath{\indexd}{j}

\safemath{\project}{P}

\newcommand{\mb}[1]{{\mathbf #1}}